%% file: ms.tex
\documentclass[apjl]{emulateapj}
\usepackage{apjfonts}
\usepackage{ulem}

\newcommand{\zsun}{$Z_\odot$}
\newcommand{\msun}{$M_\odot$}

\newcommand{\hi}{H\,{\sc i}\rm}
\newcommand{\hii}{H\,{\sc ii}\rm}
\newcommand{\hei}{He\,{\sc i}\rm}

\newcommand{\nii}{[N\,{\sc ii}]}

\newcommand{\oiii}{[O\,{\sc iii}]}
\newcommand{\oii}{[O\,{\sc ii}]}

\newcommand{\sii}{[S\,{\sc ii}]}
\newcommand{\ariii}{[Ar\,{\sc iii}]}
\newcommand{\neiii}{[Ne\,{\sc iii}]}

\newcommand{\te}{$T_e$}
\newcommand{\hbeta}{H$\beta$}
\newcommand{\halpha}{H$\alpha$}
\newcommand{\lin}{$\,\lambda$}
\newcommand{\llin}{$\,\lambda\lambda$}
\newcommand{\rhii}{$R_{\rm H\,\mbox{\tiny II}}$}
\newcommand{\rtf}{$R_{25}$}

\newcommand{\oh}{12\,+\,log(O/H)}
\newcommand{\ohtwo}{\mbox{12\,+\,log(O/H)\,=\,}}
\newcommand{\ohthree}{\mbox{12\,+\,log(O/H)\,$\simeq$\,}}
\newcommand{\ohsun}{\mbox{12\,+\,log(O/H)$_\odot\,=\,$}}
\newcommand{\rtwothree}{R$_{23}$}

\slugcomment{Accepted for publication in The Astrophysical Journal 2009 Jan 8}

\shorttitle{Extended disk of M83}
\shortauthors{Bresolin et al.}

\begin{document}

\title{The flat oxygen abundance gradient in the extended disk of M83\footnotemark[1]} 

\footnotetext[1]{Based on observations collected at the European Southern Observatory, Chile, under program 079.B-0438.}
\author{Fabio Bresolin\footnotemark[2]} \affil{Institute for Astronomy, 2680 Woodlawn
Drive, Honolulu, HI 96822, USA \\bresolin@ifa.hawaii.edu}
\footnotetext[2]{Visiting scientist, Institute of Astronomy, University of Cambridge}

\author{Emma Ryan-Weber, Robert C. Kennicutt and Quinton Goddard} \affil{Institute of Astronomy, University of Cambridge, Madingley Road, Cambridge CB3 0HA, UK \\eryan@ast.cam.ac.uk, robk@ast.cam.ac.uk, goddard@ast.cam.ac.uk}

%\author{Fabio Bresolin\footnotemark[23], \,\,Emma Ryan-Weber\footnotemark[4], Robert C. Kennicutt\footnotemark[4] \,\,and Quinton Goddard\footnotemark[4]}
%\footnotetext[2]{Institute for Astronomy, 2680 Woodlawn Drive, Honolulu, HI 96822; bresolin@ifa.hawaii.edu}
%\footnotetext[3]{Visiting scientist, Institute of Astronomy, University of Cambridge}
%\footnotetext[4]{Institute of Astronomy, University of Cambridge, Madingley Road, Cambridge CB3 0HA, UK}

\begin{abstract}
We have obtained deep multi-object optical spectra of 49 \hii\/ regions in the outer disk of the spiral galaxy M83 (=\,NGC\,5236) with the FORS2 spectrograph at the Very Large Telescope. The targets span the range in galactocentric distance between 0.64 and 2.64 times the \rtf\/ isophotal radius ($5.4-22.3$~kpc), and 31 of them 
are located at $R>$\,\rtf, thus belonging to the extreme outer  disk of the galaxy, populated by UV complexes revealed recently by the GALEX satellite. In order to derive the nebular chemical abundances, we apply several diagnostics of the oxygen abundance, including \rtwothree, \nii/\oii\/ and the \oiii\lin4363 auroral line, which was detected in four \hii\/ regions. We find that, while inwards of the optical edge the O/H ratio follows the radial gradient known from previous investigations, the outer abundance trend flattens out to an approximately constant value. The latter varies, according to the adopted diagnostic, between \ohtwo8.2 and \ohtwo8.6  (i.e.~from approximately 1/3 the solar oxygen abundance to nearly the solar value). An abrupt discontinuity in the radial oxygen abundance trend is also detected near the optical edge of the disk.
These results are tentatively linked to
 the flat gas surface density in the outskirts of the galaxy, the relatively unevolved state of the extended disk of M83, and the redistribution of chemically enriched gas following a past galaxy encounter.

\end{abstract}

 
%==========================================================================
%1 arcmin = 1.31 kpc
% - - - - - - - - - - - - - - - - - - - - - - - - - - - -
\section{Introduction}

The notion that recent or ongoing star formation activity extends well beyond the optical radii of spiral galaxies has developed
during the last decade, with the detection of both \hii\/ regions (\citealt{Ferguson:1998,Lelievre:2000}) and young B stars (\citealt{Cuillandre:2001,Davidge:2007}) out to approximately twice the 25th magnitude $B$-band isophotal radii (\rtf). Recent observations with the GALEX satellite have also traced the presence of 
massive star formation beyond the \rtf\/ radius in the form of UV-bright clusters and unstructured UV emission
(\citealt{Thilker:2005,Thilker:2007,Gil-de-Paz:2005}).
Contrary to the paucity of galaxies with \hii\/ region detections at large galactocentric distances, 
these ``extended UV'' (XUV) disks are common phenomena, being found in about 1/3 of nearby spiral galaxies (\citealt{Zaritsky:2007, Thilker:2007}). This can be interpreted in terms of the different lifetimes of O stars (responsible for the ionization of the \hii\/ regions, $\sim10$ Myr) compared to less massive B stars (producing the bulk of the UV flux, several hundred Myr). Observationally, the ages measured for the UV clusters agree with this picture, since they range between a few Myr and $\sim1$ Gyr (\citealt{Zaritsky:2007, Dong:2008}).  Alternative explanations are based on the low star formation rate (SFR) and gas density regimes 
encountered in outer spiral disks, which can affect the upper stellar mass distribution in star forming regions
(\citealt{Weidner:2005,Krumholz:2008}),
resulting  in a lack of massive star production at high galactocentric radii.

The UV data do not show  as pronounced a  drop in the SFR as the \halpha\/ surface brightness profiles do
at a radius \rhii\/ (near the optical radius), which is usually interpreted as 
the point where the surface density $\Sigma_{\rm gas}(r)$ of the gas reaches a critical threshold $\Sigma_c(r)$, below which the gaseous disks of spiral galaxies become gravitationally stable  (\citealt{Kennicutt:1989,Martin:2001}). The idea that localized density enhancements [$\Sigma_{\rm gas}(r) > \Sigma_c$(r)] in subcritical environments can produce observable star-forming sites beyond the threshold radius, as postulated by \citet{Martin:2001},
is corroborated by recent modeling that considers either spiral density waves propagating into the extended neutral gas disks (\citealt{Bush:2008}) or additional triggering mechanisms for star formation other than spontaneous gravitational instabilities (\citealt{Elmegreen:2006}). Observational support for this picture has been recently provided by \citet{Dong:2008}, who found that at the locations of the UV-bright clusters in the outer disk of M83 the gas density is near the critical value.  A different interpretation of the discrepant behavior of the \halpha\/ and UV radial profiles has been proposed more recently by  \citet{Pflamm-Altenburg:2008}, who
invoked the properties of clustered star formation and the dependence of the maximum stellar mass on the SFR density and cluster mass.

\hii\/ regions and UV knots located beyond the optical edges of spirals not only provide tracers of the star formation processes (\citealt{Thilker:2007}) and the kinematics (\citealt{Christlein:2008}) of the outer disk, 
but also offer the opportunity to explore the chemical evolution of a still poorly known component of the host galaxies. Infall models of galaxy formation predict that spiral disks build up via accretion by growing inside-out  (\citealt{Matteucci:1989, Boissier:1999}), so that the outermost parts of the disk should possess low metallicities and could possibly be still subject to active phases of infall. Consequently, observational data on galactic abundance gradients along extended radial baselines can provide strong constraints for chemical evolution models of galaxies (\citealt{Chiappini:2001,Cescutti:2007}).

The few studies of the chemical composition of \hii\/ regions at large galactocentric distances that are available in the literature suggest that the extended disks are relatively unevolved systems. In their spectroscopic work on the late-type spirals NGC~628, NGC~1058 and NGC~6946, \citet{Ferguson:1998a} found that the abundance of oxygen (a proxy for the overall metallicity in ionized nebulae) decreases in a log-linear fashion from \ohtwo8.9 near the center to \ohtwo8.0 for the outermost (1.5-2.0\,\rtf) targets [i.e.~from nearly twice the solar abundance down to 1/5 solar, if \ohsun8.66, the value by \citealt{Asplund:2005}, is adopted]. The N/O ratio they measured in the outermost regions is approximately 20-25\% of the solar value. The major limitation of this early study is represented by the fact that in the three target galaxies only a total of nine \hii\/ regions located at $R>$\,\rtf\/ were analyzed, so that general conclusions about the shape of the radial abundance gradients and the scatter in metallicity 
become very uncertain. The low oxygen abundances measured in outer disks by \citet{Ferguson:1998a}  agree, however, with the values obtained near the optical edges of a few spirals (e.g.~M101: \citealt{Garnett:1994,van-Zee:1998a,Kennicutt:2003}). More recently, \citet[=~G07]{Gil-de-Paz:2007} have reported similarly low metallicities (10-20\% of the solar value for oxygen) for \hii\/ regions in the extended disks of M83 and NGC~4625, but with a high N/O ratio, close to solar, in the case of M83.

Measuring chemical abundances in extended spiral disks is technically challenging. The typical \halpha\/ luminosity 
of the outermost targets investigated by \citet{Ferguson:1998a} and \citet{Lelievre:2000} is $L(\rm H\alpha) < 10^{38}$\,erg\,s$^{-1}$, which corresponds to the ionizing output of a small number of O stars.  These luminosities translate into extremely faint fluxes in the emission lines that are used as metallicity diagnostics. In particular, the detection of the auroral lines (such as \oiii\lin4363)
used in the classical chemical abundance analysis of ionized nebulae to derive the electron temperature (e.g.~\citealt{Pagel:1997, Stasinska:2007}) would become extremely difficult in all but the brightest outer disk \hii\/ regions in galaxies at distances of just a few Mpc. Statistical indicators of gas abundances, based on stronger nebular lines,
can be employed as an alternative. For instance, taking advantage of the presence of both singly and doubly ionized oxygen line transitions in the optical spectra of \hii\/ regions, \citet{Pagel:1979} introduced the metallicity indicator R$_{23}$\,=\,(\oii\lin3727 + \oiii\llin4959, 5007)/\hbeta\/ (both \citealt{Ferguson:1998a} and G07 base their outer disk chemical abundances on \rtwothree). As already explained in detail in a number of papers, the calibration of these indexes in terms of oxygen abundance is problematic, and can lead to still poorly understood systematic discrepancies (\citealt{Perez-Montero:2005, Bresolin:2006, Kewley:2008}). 

In this work we look at the chemical abundance properties, in particular the radial oxygen abundance gradient and the N/O abundance ratio, of the outer disk of M83, the prototypical XUV disk galaxy (\citealt{Thilker:2005}). While the number of UV complexes with associated \hii\/ emission is only a fraction of the total of more than 100, there are still several dozen faint \hii\/ region candidates beyond the optical edge of the galaxy that are visible in deeply exposed \halpha\/ frames. Within the context of a project that aims at obtaining 
chemical abundance properties in extended spiral disks, we have targeted these emission-line objects with a multi-slit spectroscopic program. While a similar project was published by G07 at the time we started to acquire our data, we overcome some of the deficiencies encountered in their  work,  by enlarging the sample of outer disk \hii\/ regions ($>30$ targets at $R>$\,\rtf), and  by obtaining deeper spectra. In order to facilitate the comparison with inner disk nebular abundances in M83 (\citealt{Bresolin:2002, Bresolin:2005}), we have also included in our sample a number of \hii\/ regions that are located inside the \rhii\/ radius (taken from \citealt{Martin:2001}: 5\farcm1 = 6.7~kpc at the adopted distance of 4.5~Mpc, \citealt{Thim:2003}). Our principal goal at the onset of the project was to verify whether in the prototypical XUV disk galaxy the oxygen abundance decreases steadily with galactocentric distance, down to values comparable to those found in previous investigations (10-20\% of the solar value). 

Here we present the results of our study of the outer disk of M83.
The observational material and the analysis of the multi-object spectroscopy are presented in \S2. In \S3 we carry out the chemical composition analysis, using a variety of metallicity indicators. Our discussion concerning the observed oxygen abundance gradient in M83, from the inner part of the disk to the outermost \hii\/ region (at 2.6\,\rtf), and the interpretation of the results, are presented in \S4. Lastly, we summarize our findings in \S5.

% - - - - - - - - - - - - - - - - - - - - - - - - - - - -
\section{Observations and data reduction}

Candidate \hii\/ regions in the outer disk of M83 were selected from deep H$\alpha$ images obtained at the 90-inch Bok telescope of the Steward Observatory with the 90Prime camera. Several of these \hii\/ regions have a counterpart in the list of UV-emitting clusters  detected by GALEX (\citealt{Thilker:2005}; G07). 
The azimuthal distribution of the candidates is not uniform, since, as shown in the GALEX images, the brightest star-forming sites in the outer disk of M83 are organized in discrete filamentary structures that extend outwards from the edge of the optical disk of the galaxy, in general corresponding with enhancements in the
radio \hi\/ emission.

\begin{figure}
\medskip
\center \includegraphics[width=0.475\textwidth]{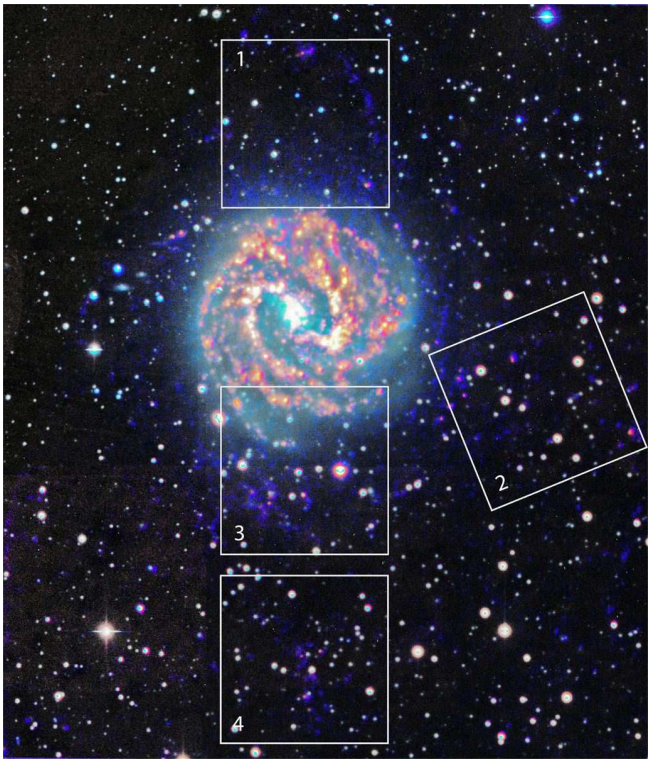}\medskip
\caption{Distribution of the four fields in which the multi-object spectroscopy has been carried out. Each field is  6\farcm8 $\times$ 6\farcm8 wide. The image is a composite of GALEX and \halpha\/ data, with UV-bright knots appearing in blue, and \hii\/ regions in red.\\
 \label{image}}
\epsscale{1}
\end{figure}

Optical spectra were acquired with FORS2 at the Very Large Telescope of the European Southern Observatory on Paranal. Objects distributed in four non-overlapping 6\farcm8 $\times$ 6\farcm8 fields were observed  in service mode between May 10, 2007 and June 19, 2007. One of these fields covers part of the northern extended disk of the galaxy, where 
a faint UV filament, resembling an external spiral arm, extends to the north of  the main disk of the galaxy.
Two fields are located to the south (one of them partially overlapping with the bright optical disk), while the last field is located to the south-west of the optical disk (Fig.~\ref{image}).
The slits of the multi-object masks used for the observations included, in addition to some prominent UV-emitting knots,  a number of additional emission-line objects identified on H$\alpha$ pre-imaging frames. In filling the slit masks priority was given to the outermost \hii\/ region candidates.
The spectra were obtained through 1.0 arcsec-wide slitlets with the 600B grism (`blue' setting, central wavelength 4650\,\AA, 6\,$\times$\,970\,s exposures) and with the 1200R grism (`red' setting, central wavelength 6500\,\AA, 3\,$\times$\,335\,s exposures), providing an almost full spectral coverage from approximately 3500\,\AA\/ to 7200\,\AA\/ for most of the targets (the exact spectral coverage depending on the location of the objects within the field). With these setups, the FWHM spectral resolution is 4.5\,\AA\/ and 2.4\,\AA\/ for the blue and red spectra, respectively. The red and blue observations were carried out in consecutive exposures, thus minimizing possible differences in target coverage arising from variations in the seeing. Spectra for the southern field were not obtained in the same night, but the image quality during the observations was similar.
Typical seeing conditions during the observations were $\sim 0.8$ arcsec. The use of the FORS atmospheric dispersion corrector (\citealt{Avila:1997}) allowed us to avoid the adverse effects of
differential refraction on the emission line ratios and the derived chemical abundances.
Observations of the standard star LTT~4816 have been obtained to flux-calibrate the \hii\/ region spectra.

For the data reduction we relied on the flat-fielded and wavelength calibrated products of the EsoRex pipeline (version 3.1.1). We then obtained flux-calibrated spectra, clean of cosmic rays, using common {\sc iraf}\footnote{{\sc iraf} is distributed by the National Optical Astronomy
Observatories, which are operated by the Association of Universities
for Research in Astronomy, Inc., under cooperative agreement with the
National Science Foundation.} routines. Since the standard star spectra were obtained through a single slit near the center of the detector, we investigated possible variations in the response function across the CCD by comparing 
the wavelength dependence of flat fields obtained from slitlets distributed across the FORS2 field. Once scaled to a common mean flux, the resulting rms dispersion of the flat field spectra is $\sim$\,0.03\,mag, which is also our adopted uncertainty for the flux calibration.

\input{tab1}

Emission-line fluxes were measured for 49 \hii\/ regions belonging to M83. Their coordinates are 
presented in Table~\ref{sample}, together with the galactocentric distances $R$/\rtf\/ (adopting \rtf\,=\,6.44 arcmin from \citealt{de-Vaucouleurs:1991}, an inclination angle  $i=24^\circ$ from \citealt{Heidmann:1972}, and a
value for the position angle of the major axis $\theta=45^\circ$ from \citealt{Comte:1981}; we did not account for the warp of the disk at large radii, \citealt{Rogstad:1974}), and the identification of the objects in common with the study by G07. As Table~\ref{sample} shows, our sample covers \hii\/ regions from the outer third of the main disk of star formation of M83, out to 2.6 times the isophotal radius (\rtf).
A small number of targets, although detected in the H$\alpha$ line, were too faint to allow a measurement of additional lines in their spectra, and were therefore discarded from the subsequent analysis. We also removed from the final sample a number of background line-emitting galaxies (listed in the Appendix).

For most of our targets we were able to measure the intensities of the strongest recombination lines of the hydrogen Balmer series, and the collisionally excited metal lines \oii\lin3727, \oiii\llin4959, 5007, \nii\llin6548, 6583, and \sii\llin6717, 6731. In a handful of cases the signal-to-noise ratio of the spectra was sufficient to detect some of the fainter metal lines, such as \neiii\lin3869 and \ariii\lin7135. In four \hii\/ regions we also measured the \oiii\lin4363 auroral line, which, in combination with \oiii\llin4959, 5007, provides a direct determination of the electron temperature of the ionized gas (as described in \S~\ref{abundances}). Due to the moderate redshift of the galaxy (v\,=\,$513$ km\,s$^{-1}$, \citealt{Koribalski:2004}), the measurement of the strongest helium recombination line, \hei\lin5876, was hampered by the contamination of the sodium telluric D$_2$ line at 5890~\AA. 
The other helium lines were generally too faint to be measurable.

As a rule, the blue and red spectra have no emission lines in common. We then decided to refer the strengths of the emission lines in the blue wavelength range (in particular the \oii\/ and \oiii\/ lines)
to the strength of H$\beta$, while in the red we express the line strengths of \nii\/ and \sii\/ relative to H$\alpha$, and assume the case B ratio H$\alpha$/H$\beta$\,=\,2.86 (valid at T$_e$\,=\,10$^4$\,K) in order to match the two sets of lines (this renormalization was done after the correction for reddening was applied to the 
individual line fluxes). 
Since, with the exception of \oii\lin3727, all the key forbidden lines used for the metallicity determination are close
to their reference Balmer lines, the diagnostic ratios are fairly insensitive to reddening. 
The interstellar reddening was determined from the observed H$\gamma$/H$\beta$ ratio (these two Balmer lines have been detected in all 49 \hii\/ regions) and, when allowed by the signal-to-noise ratio, the H$\delta$/H$\beta$ ratio, compared to the theoretical values for case B at 10$^4$\,K. The resulting reddening-corrected line fluxes, expressed in units of H$\beta$\,=\,100, are presented in Table~\ref{fluxes}. For the \oiii\/ and \nii\/ doublets we provide the fluxes of the strongest line of the pair (\lin5007 and \lin6583, respectively). For \sii\/ we provide the sum of the \lin6717 and \lin6731 fluxes.
The data for \oii\lin3727 and \ariii\lin7135 are sometimes missing due to incomplete wavelength coverage.
The errors quoted in Table~\ref{fluxes} include, besides the statistical errors and an estimate of the flux-calibration errors, also the uncertainty in the reddening correction, which could be significant in particular for the \oii\lin3727 line. In general, the extinction values  (shown in column 8) are small, 
with an average c(H$\beta$)\,=\,0.08 for \rtf\,$>$\,1, compared to an average c(H$\beta$)\,=\,0.17 for 
\rtf\,$<$\,1.

\input{tab2}

\begin{figure}
\medskip
\center \includegraphics[width=0.475\textwidth]{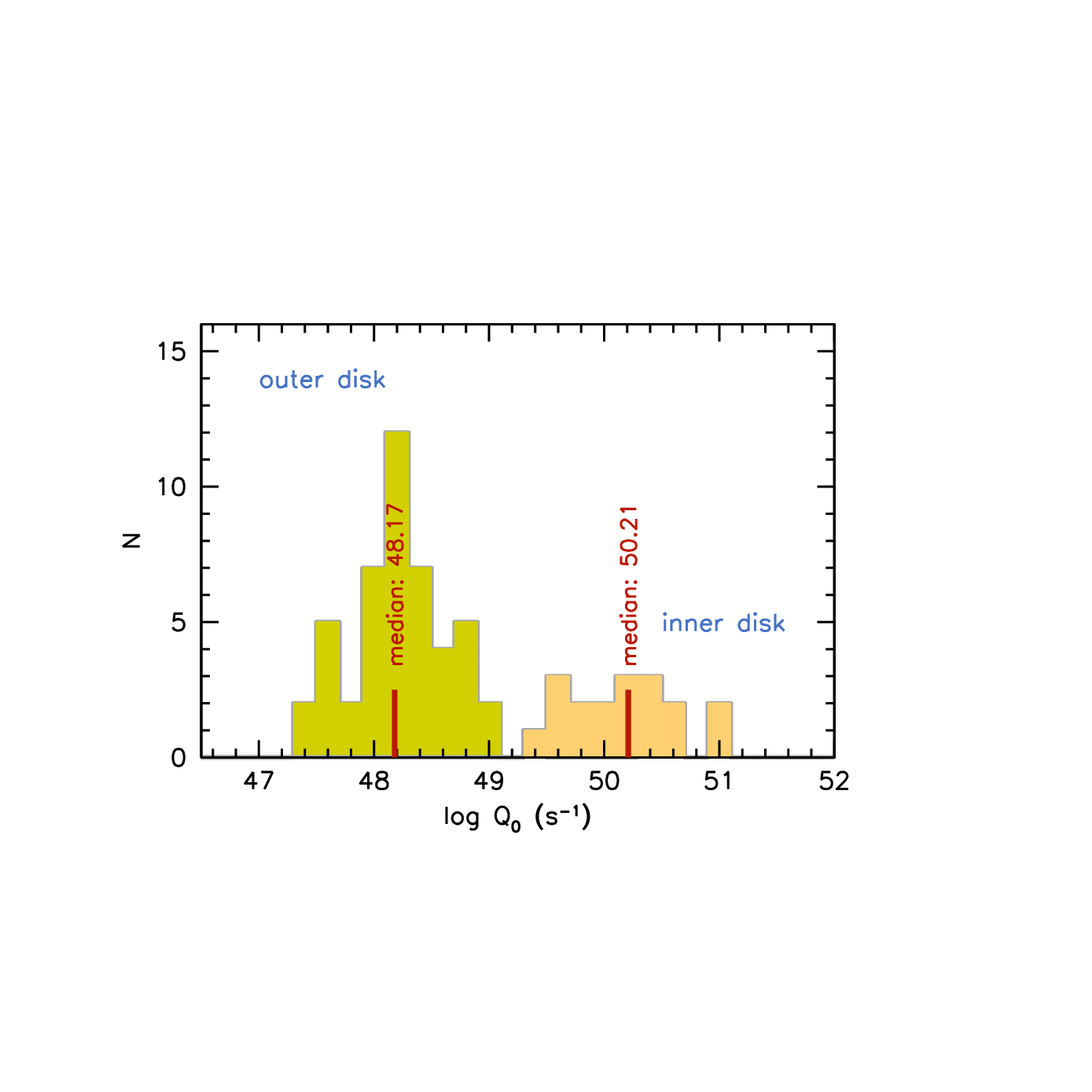}\medskip
\caption{Histograms showing the distribution of the extinction-corrected number of ionizing photons Q$_0$ (in s$^{-1}$) for the 
\hii\/ regions in the outer disk of M83 (green shaded area, left) and for the samples of inner disk regions from  
\citet{Bresolin:2002} and \citet[yellow shaded area, right]{Bresolin:2005}. The median values of the two distributions
are indicated by the red marks.\\ \\
 \label{qo}}
\end{figure}

The H$\beta$ line flux in column (7) is the flux measured within the 1-arcsec slits, corrected for extinction, and likely underestimates the total nebular flux, due to slit slosses.
As already pointed out by G07, these faint nebular fluxes correspond to the ionizing output of one late-O star or, at most, a few massive stars. Fig.~\ref{qo} compares the histogram of the extinction-corrected number of ionizing photons Q$_0$ corresponding to the H$\beta$ fluxes in Table~\ref{fluxes} (for a distance of 4.5~Mpc) to that obtained from two samples of inner disk \hii\/ regions studied by \citet{Bresolin:2002} and \citet{Bresolin:2005}. As the histograms show, the median value of our outer disk sample is log\,Q$_0$\,=\,48.17 (approximately equivalent to the ionizing output of a single O8.5\,V star, \citealt{Martins:2005}), while in the giant \hii\/ regions of the inner disk the median value is two orders of magnitude larger, log\,Q$_0$\,=\,50.21 ($\sim$130 O8.5\,V stars, or four O3\,V stars). It should be noted that the aperture corrections are likely to be larger for the supergiant \hii\/  regions that make up the inner disk sample compared to the more compact outer disk regions, which would increase the offset seen in Fig.~\ref{qo} between the two samples.

\begin{figure}
\medskip
\center \includegraphics[width=0.475\textwidth]{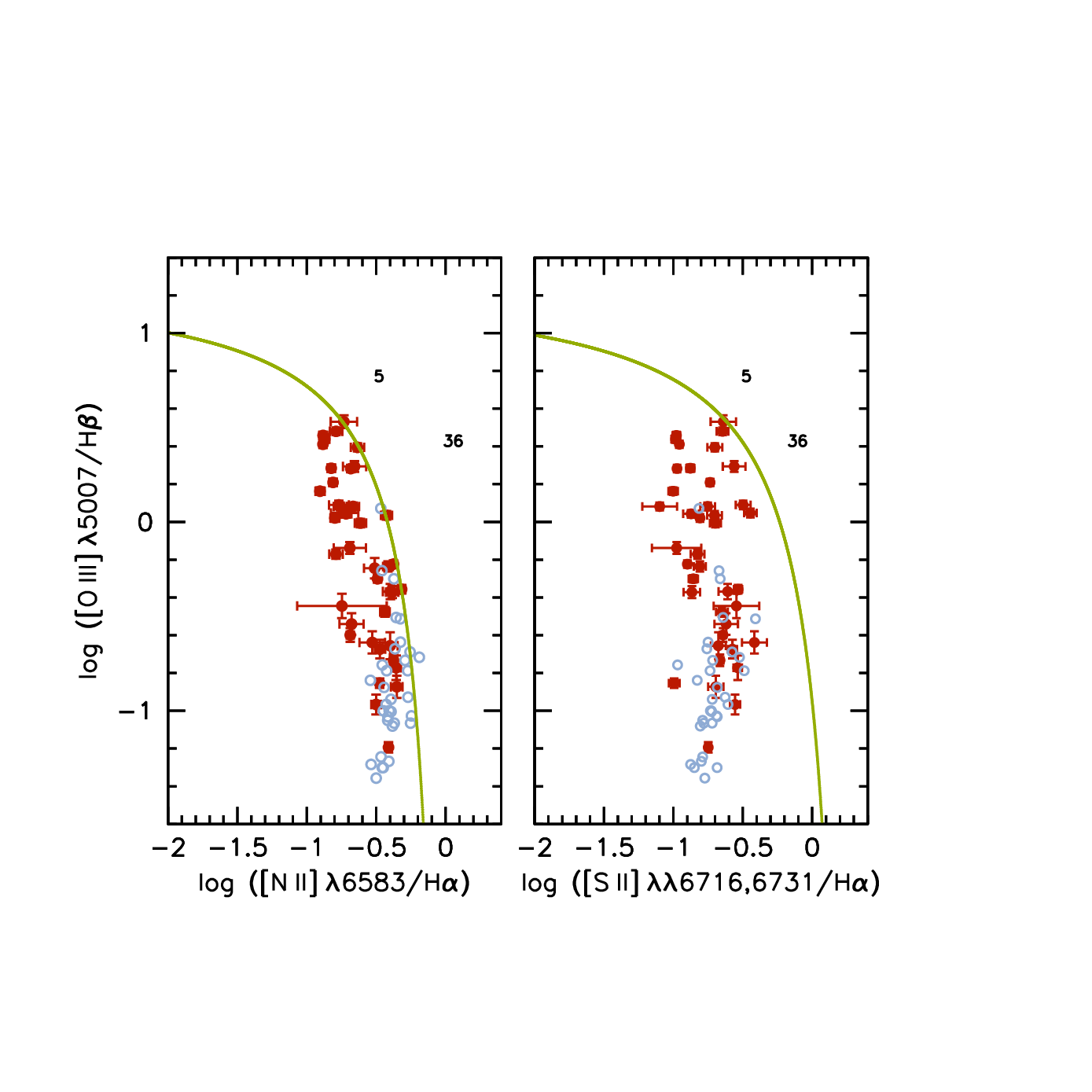}\medskip
\caption{Emission-line diagnostic diagrams for the M83 outer disk sample (red dots) and for the inner disk \hii\/ regions of \citet{Bresolin:2002} and \citet[open light blue circles]{Bresolin:2005}. The curves represent the upper boundaries for photoionized nebulae defined by \citet{Kewley:2006a}. The two outlying regions, \#5 and \#36, are identified.\\
 \label{bpt}}
\end{figure}

A quick characterization of the \hii\/ regions in the outer disk of M83 can be obtained by plotting the 
traditional \nii/\halpha\/ vs.~\oiii/\hbeta\/  and \sii/\halpha\/ vs.~\oiii/\hbeta\/ diagnostic diagrams (\citealt{Baldwin:1981,Veilleux:1987}). Fig.~\ref{bpt} shows that the excitation sequence formed by the outer disk \hii\/ regions merges into the sequence outlined by the inner disk \hii\/ regions, which are found at lower excitation (smaller \oiii/\hbeta\/ ratio), likely as a result of the higher metal content near the center of the galaxy. Two objects, number 5 and 36 in our list, deviate from the main sequences of points in Fig.~\ref{bpt}, and 
lie above the curves that define the upper boundaries of the loci occupied by star-forming 
regions (taken from \citealt{Kewley:2006a}). The ionization structure of these two \hii\/ regions appears to differ from that of the remaining objects, and in particular the
high \nii/\halpha\/ and \sii/\halpha\/ ratios measured in object 36 resemble the signatures of shock heating typically observed in supernova remnants. This object was in fact already identified as a supernova remnant candidate on the basis of the strong \sii/\halpha\/ ratio by \citet[their object BL34]{Blair:2004}. Its emission line ratios
are consistent with this conclusion if one compares the location of supernova remnants and photoionized nebulae in a \sii/\halpha\/ {\it vs.} \nii/\halpha\/ plot (\citealt{Sabbadin:1977}). We will remove these two targets (5 and 36) from the remainder of the discussion and from all of the following diagrams.

We have compared our fluxes in Table~\ref{fluxes} with the values published by G07 for the eight targets in common.
The agreement is satisfactory for the \nii, \sii\/ and \oiii\/ fluxes in most cases. An obvious discrepancy
is represented by our target \#30 (XUV~02 in G07), with our \oiii\lin 5007 flux being more than $6\times$ smaller. 
Although the coordinates of our target match, within the uncertainties, those by G07, we believe that the 
slits did not actually include the same object or the same portion of it. In all cases we measure stronger \oii\lin 3727 fluxes (5 of the 8 objects in common have a 3727 flux in G07), despite the fact that we derive
smaller extinction values (0.06 vs. 0.16 for the average c(H$\beta$) of the objects in common). The worst case (factor of 4 discrepancy) is represented by \#42 (XUV~22 in G07). For this target the small signal-to-noise ratio in the G07 spectrum is likely the reason. It is well known that \oii\lin3727 line fluxes are the most uncertain in nebular spectra, being the most affected by errors in the flux calibration and in the interstellar reddening correction. While it is difficult to assess the accuracy of the line ratios, we have verified in the few cases were this was possible that 
the strengths  of high order Balmer lines (H9, H10 and H11) relative to \hbeta\/ agree with the theoretical Balmer decrement. This provides  some confidence about the reliability of our fluxes for the nearby \oii\lin3727 line.

% - - - - - - - - - - - - - - - - - - - - - - - - - - - -
\section{Nebular abundances}\label{abundances}

\subsection{Electron temperatures and direct abundances}\label{direct}

In order to derive reliable chemical abundances of ionized nebulae it is necessary to have a good knowledge of the physical conditions of the gas, in particular of the electron temperature \te, because the line emissivities
depend strongly on it. Nebular electron temperatures can be obtained from lines of the same ions that originate from different excitation levels, such as the auroral \oiii\lin4363 and the nebular \oiii\llin4959, 5007 lines.
In the case of high-excitation (low metallicity) extragalactic \hii\/ regions and planetary nebulae the \oiii\lin4363 line is often detected, but it becomes unobservable as the cooling of the gas becomes efficient at high metallicity, or whenever the objects are faint. 

Because of the intrinsically weak levels of line emission from the \hii\/ region population in the outer disk of M83, in which the ionizing sources appear to be single stars or very small clusters, it is not surprising that the \lin4363 line remains undetected for the vast majority of the objects. 
The chemical abundance analysis must then be based on statistical methods, without knowledge of the electron temperature (as discussed in the next section).
However, we obtained significant \oiii\lin4363 detections in four  \hii\/ regions, which allowed us to derive their oxygen abundance via the direct method based on the knowledge of \te. In Table~\ref{auroral} we present their identification and the measured \lin4363 fluxes. Using the routines in the {\tt nebular} package of {\sc iraf} we have derived the electron temperature (column 3), the total oxygen abundance \oh\/ (column 4), and the N/O abundance ratio (column 5), assuming that N/O\,=\,N$^+$/O$^+$. 

While we postpone a full discussion of the chemical compositions to the next section, we briefly note that: {\it (a)} one of the outermost \hii\/ regions in our sample, \#20 at $R$/\rtf\,=\,2.25, has an abundance \ohtwo8.17, equivalent to 1/3 of the solar value. Two of the remaining objects, \#39 and \#43, located between 1.25 and 1.30 isophotal radii, have similar oxygen abundances, while the remaining \#29 appears to be even more metal-rich (half-solar abundance), despite being located well beyond the isophotal radius at $R$/\rtf\,=\,1.21. The results from this limited number of \lin4363 detections suggest that the outer disk of M83 is, perhaps somewhat surprisingly, not as metal-poor as expected. G07 assigned an abundance \ohtwo7.86 to region \#20 (their NGC~5236:\,XUV\,01) based on the value of the abundance indicator \rtwothree. Our metallicity is a factor of two larger. {\it (b)} the N/O ratio ranges approximately between 50\% and 80\% of the solar value [log(N/O)$_\odot$\,=\,$-0.88$, \citealt{Asplund:2005}]. As we will show in the next section, these N/O ratios are similar to those found in extragalactic \hii\/ regions of comparable O/H abundances.

%\clearpage
\input{tab3}

\subsection{Strong-line abundances}
In the absence of direct indicators of the electron temperature such as the auroral lines, the chemical analysis must be carried out via methods that statistically determine an oxygen abundance from the ratios of strong nebular lines.
These strong-line indicators and their calibration in terms of oxygen abundance have been discussed at length in recent years, due especially to the necessity of resorting to these methods for the measurement of abundances in star-forming galaxies at intermediate and high redshift. In order to obtain a picture of the chemical composition in the outer disk of M83, we will consider a number of strong-line indicators. It is  well-known that different indicators provide different solutions to the chemical composition of star-forming galaxies and \hii\/ regions. Some of the methods only use oxygen lines to estimate oxygen abundances, while others use lines of other elements (e.g.~nitrogen) as proxies of the oxygen abundance. Moreover, the same strong-line indicator can be calibrated empirically from nebulae where a \te-based abundance is available, or from grids of photoionization models. Once again, systematic differences are found, with models providing higher abundances (up to 0.7 dex) than \te-based methods (\citealt{Kewley:2008}). Some indicators are strongly dependent on the ionization parameter, while others are almost insensitive to it.
By considering different strong-line indicators we attempt to provide robust results concerning the trends of chemical composition in the extended disk of M83, with the warning that a significantly more detailed picture is beyond 
our current capabilities. 
One potential difficulty in the use of strong-line indicators for our sample of low-luminosity \hii\/ regions is that the available calibrations are based 
on more luminous giant and supergiant \hii\/ regions. However, tests of strong-line diagnostics in Galactic nebulae ionized by one or a few stars show no evidence for systematic effects (\citealt{Kennicutt:2000, Oey:2000}). As a further empirical test, we have applied one of the diagnostics described below (\nii/\oii) to a number of well-studied \hii\/ regions in the Milky Way, including the Orion nebula (\citealt{Esteban:2004}), M8, M17 and others (see \citealt{Garcia-Rojas:2007} and references therein).
Admittedly, for the sample considered  \oh\/ only ranges between 8.39 and 8.56. 
With the particular \nii/\oii\/ calibration adopted (\citealt{Bresolin:2007}, obtained from extragalactic giant \hii\/ regions) we were able to recover the published T$_e$-based O/H abundances within 0.12 dex in all cases, with a mean difference of 0.04 dex and a standard deviation of 0.05 dex. The smallness of the latter value indicates that the adopted strong-line diagnostic is quite reliable also when applied to lower luminosity nebulae, at least near the O/H range of the Galactic sample.
\smallskip

The strong-line indicators considered in this work are: \smallskip

\noindent
{\it (a)} R$_{23}$\,=\,(\oii\lin3727 + \oiii\llin4959,5007)/\hbeta. Despite its drawbacks (e.g.~\citealt{Perez-Montero:2005}), this classic indicator is widely used. One of the main difficulties in its adoption is related to the necessity of locating  which of two branches (upper and lower) a given \hii\/ region belongs to, since R$_{23}$ is degenerate. This is usually accomplished by simultaneously looking at line ratios that increase monotonically with oxygen abundance, such as \nii/\halpha\/ or \nii/\oii. According to \citet{Kewley:2008}, an object belongs to the upper branch of R$_{23}$ if 
log(\nii\lin6583/\oii\lin3727)~$>$~$-1.2$ and/or log(\nii\lin6583/\halpha)~$>$~$-1.1$ [the lowest values in our sample are log(\nii/\oii)\,=\,$-1.1$, log(\nii/\halpha)\,=\,$-0.90$]. According to this scheme, all of the \hii\/ regions in our sample belong to the R$_{23}$ upper branch [\ohtwo$>8.4$ in the abundance scale calibrated via photoionization models].  For the derivation of oxygen abundances we adopt the upper branch calibration of \citet{McGaugh:1991}, in the analytical form given by \citet{Kobulnicky:1999}.

\smallskip
\noindent
{\it (b)} N2\,=\,\nii\lin6583/\halpha, as empirically calibrated by \citet{Pettini:2004}. This indicator, like the remaining ones considered by us, is a monotonic function of oxygen abundance, thus avoiding the degeneracy problem of R$_{23}$.

\smallskip
\noindent
{\it (c)} O3N2\,=\,log[(\oiii\lin5007/\hbeta)/(\nii\lin6583/\halpha)], also calibrated by \citet{Pettini:2004}.

\smallskip
\noindent
{\it (d)} \nii\lin6583/\oii\lin3727, which, as shown by \citet{Dopita:2000}, is highly insensitive to the ionization parameter. We adopt here two different calibrations, the theoretical one by \citet{Kewley:2002}, based on grids of photoionization models, and the empirical one by \citet{Bresolin:2007}, based on a sample of
\hii\/ regions with available \te-based abundances. 

\smallskip
\noindent
{\it (e)} Ar3O3\,=\,log(\ariii\lin7135/\oiii\lin5007), empirically calibrated by \citet{Stasinska:2006}. Although we measured the \ariii\lin7135 line only in a minority of cases, we considered this indicator, monotonic as a function of O/H, because it does not involve the use of the \nii\lin6583 line. G07 proposed that the N/O ratio of \hii\/ regions in the extended disk of M83 could be higher than normal for their oxygen abundance, which would affect the oxygen abundances obtained from N2, O3N2 and \nii/\oii. While the results obtained from the auroral lines (Section~\ref{direct}) do not support this idea, the Ar3O3 indicator could reveal obvious problems in abundances obtained from line ratios involving \nii, and in the choice of upper/lower branch of R$_{23}$.

The results of the application of these strong-line indicators to our \hii\/ region sample are shown in Fig.~\ref{abundances}, where we plot O/H abundances as a function of galactocentric distance. In this figure we also include comparison data for the inner disk of M83 from \citet[open light blue circles]{Bresolin:2002} and \citet[light blue squares]{Bresolin:2005}. The vertical dotted line in the diagrams represents the threshold radius \rhii\/ (5\farcm1 = 6.7~kpc) defined by \citet{Martin:2001} as the position where a strong break in the azimuthally averaged \halpha\/ surface brightness profile is encountered. 
Our first remark looking at the abundance trends in Fig.~\ref{abundances}
is that the new \hii\/ region data merge nicely with the comparison sample data. This is true also for the diagnostic diagrams in Fig.~\ref{bpt}, thus excluding significant systematic offsets in our line fluxes.

Qualitatively, the plots in Fig.~\ref{abundances} look similar to each other, although the scatter of the points varies for different indicators, being smaller in the case of \nii/\oii. The few datapoints available in the case of Ar3O3 (panel $f$) are in rough agreement with other indicators, suggesting that the choice of the upper branch of R$_{23}$ (panel $c$), made on the basis of the strength of the \nii\lin6583 line relative to \halpha\/ and \oii\lin3727, is correct throughout the outer disk sample, and also confirms what is apparent from all the different indicators, i.e.~that the oxygen abundances in the extended disk are not as low as 10\%-15\% of the solar value, even though the absolute values really depend on the particular choice of abundance diagnostic. Concentrating now on the remaining diagrams, the scatter of the points is largest in the case of the N2 (panel $a$) and O3N2 (panel $b$) indicators. Looking at the inner disk data, both diagnostics seem to be affected by a saturation effect at high abundance. This is a known effect (\citealt{Pettini:2004, Kewley:2008}). A log-linear upward trend in O/H towards the center of the galaxy is instead very clear in the case of the \nii/\oii\/ plot. The familiar systematic offset between calibrations of abundance diagnostics based on photoionization model grids and empirical measurements is obvious in the comparison between panel $d$ (\nii/\oii\/ indicator calibrated from the models by \citealt{Kewley:2002}) and panel $e$ (same indicator, calibrated empirically by \citealt{Bresolin:2007}). In this case, the empirical method provides oxygen abundances that are nearly 0.5 dex below the theoretical method. Higher abundances are also provided by the theoretical calibration of R$_{23}$ (panel $c$) compared to the empirical calibrations.
The radial trends are, however, virtually identical. 

\begin{figure*}
\medskip
\center \includegraphics[width=0.9\textwidth]{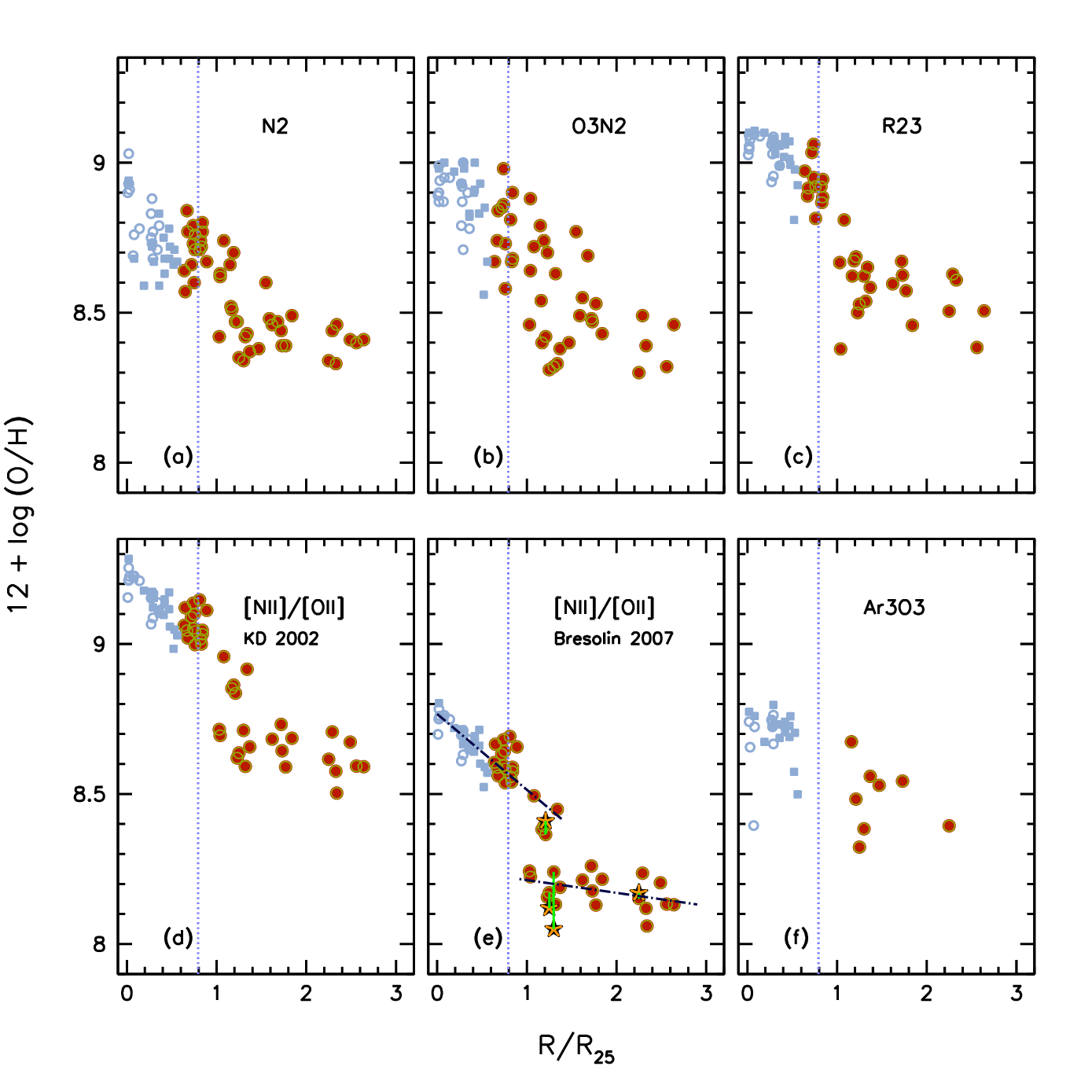}\medskip
\caption{Radial abundance gradient in M83, where the oxygen abundance has been determined from different strong-line methods: (a) N2; (b) O3N2; (c) R$_{23}$; (d) \nii/\oii\/ and the theoretical calibration by \citet{Kewley:2002}; (e) \nii/\oii\/ and the empirical calibration by \citet{Bresolin:2007}; (f) Ar3O3. The new M83 sample is represented by the red dots. The open circles and the full light blue squares represent the 
inner disk regions from  
\citet{Bresolin:2002} and \citet{Bresolin:2005}, respectively. The blue dotted line at $R$/\rtf\,=\,0.79 shows the distance at which the azimuthally averaged \halpha\/ surface brightness has a sudden drop (\citealt{Martin:2001}). In panel $(e)$ we show with star symbols the four \hii\/ regions for which we have detected the \oiii\lin4363 auroral line (a line connects the \te-based abundances to the empirical ones). The linear regressions to the inner and the outer abundances (Eq.~1 and 2) are also shown with the dot-dashed lines in $(e)$.\\ \\
\label{abundances}}
\end{figure*}

The abundance offset between 
the N2 (panel $a$) and the empirical \nii/\oii\/ (panel $e$) diagnostics is puzzling at first, since both methods have been calibrated using samples of ionized nebulae for which the abundances were derived from the knowledge of the electron temperature (apart from a few objects at high metallicity used in the calibration of N2 by \citealt{Pettini:2004}). To investigate the origin of this systematic difference we have considered the \hii\/ region sample used by \citet{Bresolin:2007} to obtain his simple empirical relation between O/H and the
\nii/\oii\/ line ratio, and verified that also in this case the \citet{Pettini:2004} N2 calibration would provide abundances that are systematically $\sim$0.2 dex higher. 
This might have to do with the inclusion of \hii\/  galaxies in the \citet{Pettini:2004} calibrating sample, and with the fact that a non-negligible diffuse \nii-emitting component
would affect the integrated spectra of galaxies in such a way as to lead to a slight overestimate of the metallicity (\citealt{Moustakas:2006}). On the other hand, the simple analytical form of the empirical calibration provided
by \citet{Bresolin:2007} could be slightly underestimating the O/H ratio in the range of interest here.
Once again, however, we note that the general qualitative trends provided by different indicators are very similar. 

In panel $e$ we have included the abundances derived from the \oiii\lin4363 auroral line, described in \S~\ref{direct} for the four objects in Table~\ref{auroral}. The good agreement between the direct abundances (star symbols) and the strong-line method abundances indicates that the empirical calibration of \nii/\oii\/ by \citet{Bresolin:2007} is more tightly matched to the direct, \te-based method than the remaining diagnostics, at least for the objects considered here.

In all the diagrams of Fig.~\ref{abundances} a break in the abundance gradient is clearly seen (except for Ar3O3, for which we do not have a sufficient number of points). Despite differences in the scatter in O/H and in the absolute value of the oxygen abundance, the gradient becomes virtually flat outside the isophotal radius, or slightly outside the threshold radius \rhii\/ (blue dotted line).
The effect is more clearly visible in the case of the abundances determined with the \nii/\oii\/ line ratio, because of the smaller scatter. In fact, the diagrams in panels $d$ and $e$  suggest the presence of two distinct 
groups of \hii\/ regions, one located inwards of $R$/\rtf\,=\,1.2, for which there is a relatively steep decrease of O/H with radius, and a second 
one from the isophotal radius outwards, for which the abundance is almost constant or very slowly declining with radius.
In addition, a significant drop in metallicity occurs around $R$/\rtf\,=\,1.2.
Depending on the method, the oxygen abundance in the extended disk of M83 can be as high as \ohtwo8.6 (approximately solar, from R$_{23}$ and the theoretical calibration of \nii/\oii) or as low as \ohtwo8.2 (1/3 solar, from the empirical \nii/\oii\/ calibration). The auroral line-based abundances are in agreement with the latter value. On the other hand, a solar-like metallicity at large galactocentric distances appears difficult to reconcile with the idea that the outer fringes of spiral disks are chemically unevolved. Finally, we note that a step-like gradient in the outer disk of M83 is suggested by the abundances published by G07 (their Fig.~7a), although the absolute abundances are smaller than ours, and virtually only one out of a total of 12 \hii\/ regions with a published metallicity (XUV~01) is located well beyond the isophotal radius, with an oxygen abundance \ohtwo7.86.

In order to quantify the abundance gradient, we have carried out a linear regression to the data points shown in Fig.~\ref{abundances}(e) (\citealt{Bresolin:2007} calibration of \nii/\oii), independently for the inner and the outer disk. We have arbitrarily included in the inner disk fit the five \hii\/ regions that lie at $R>$\,\rtf\/ and \oh\,$>$\,8.3 (all relatively close to each other south and south-west of the center), since also visually they fit very well an extension of the inner gradient.
We obtained:
% the objects are 3_02, 3_11, 3_17, 3_27 and 2_33

\medskip
\noindent
$\bullet$ inner disk (54 data points, rms\,=\,0.05):
\begin{equation}
12+\log({\rm O/H}) = 8.77 (\pm0.01) - 0.25 (\pm 0.02)\, R/R_{25}
\end{equation}

\medskip
\noindent
$\bullet$ outer disk (19 data points, rms\,=\,0.05):
\begin{equation}
12+\log({\rm O/H}) = 8.25 (\pm0.04) - 0.04 (\pm 0.02)\, R/R_{25}
\end{equation}

\noindent
The same slopes, but with a systematic offset of +0.47 dex, are obtained using the \nii/\oii\/ abundances from the 
\citet{Kewley:2002} calibration. The two regressions in (1) and (2) cross at $R$/\rtf\,=\,2.26.
The slope of the gradients correspond to $-0.030$~dex\,kpc$^{-1}$ (inner) and $-0.005$~dex\,kpc$^{-1}$ (outer). The inner abundance gradient derived here is slightly steeper than the value reported earlier ($-0.022$~dex\,kpc$^{-1}$: \citealt{Bresolin:2002, Pilyugin:2006}). The slopes obtained by G07 are very different, due to the discrepancies in the comparison with our abundance analysis (\S~\ref{comparison}), and to the assumed shape of the gradient.
Expressing the inner gradient slope in terms of the isophotal radius, our value of $-0.27$~dex/\rtf\/ is similar to the mean of $-0.23$~dex/\rtf\/ for five barred 
galaxies in \citet{Zaritsky:1994}, compared to the mean for their whole sample (barred\,+\,unbarred spirals) of 
$-0.59$~dex\,kpc$^{-1}$. We stress, however, the difficulty of measuring chemical abundance gradients in galaxies, and especially the fact that
the use in the literature of different metallicity diagnostics leads to very different results. An example is provided by the nearby galaxy M33, for which the estimates of the slope of the oxygen abundance gradient
vary between $-0.012$~dex\,kpc$^{-1}$ (\citealt{Crockett:2006}) and $-0.10$~dex\,kpc$^{-1}$ (\citealt{Vilchez:1988}, see \citealt{Rosolowsky:2008}).

\begin{figure}
\medskip
\center \includegraphics[width=0.475\textwidth]{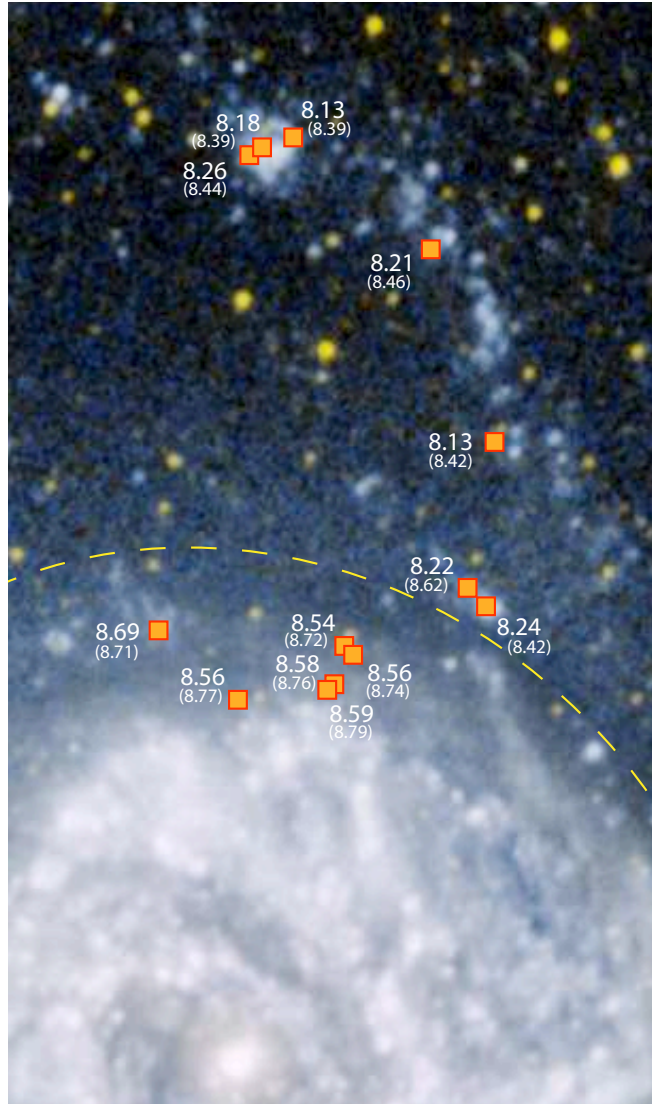}
\caption{The positions of the northern field \hii\/ regions shown as  squares on the GALEX image of M83 (courtesy NASA/JPL-Caltech), and labeled with their \oh\/ values, obtained from \nii/\oii\/ in the \citet{Bresolin:2007} scale, and from N2 in the \citet{Pettini:2004} calibration (in brackets). The nucleus of the galaxy appears at the bottom. The dashed curve represents the \rtf\/ radius.\\ 
\label{galexn}}
\end{figure}

\begin{figure}
\medskip
\center \includegraphics[width=0.475\textwidth]{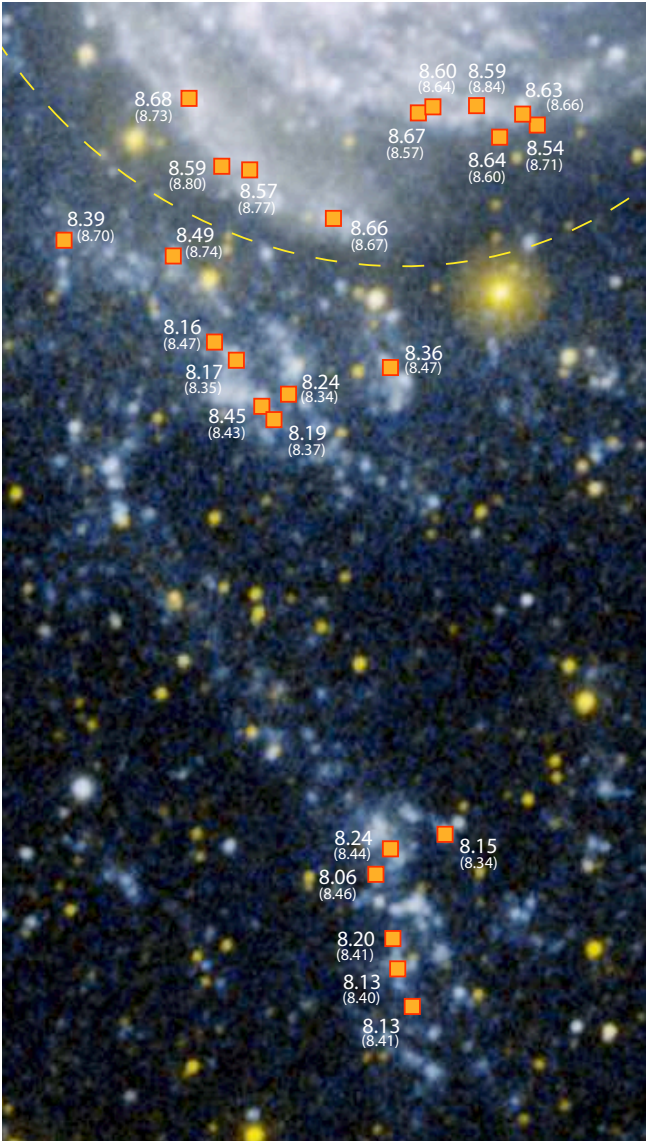}
\caption{The positions of the southern field \hii\/ regions  shown as  squares on the GALEX image of M83 (courtesy NASA/JPL-Caltech), and labeled with their \oh\/ values, obtained from \nii/\oii\/  in the \citet{Bresolin:2007} scale, and from N2 in the \citet{Pettini:2004} calibration (in brackets). The dashed curve represents the \rtf\/ radius.\\
\label{galexs}}
\end{figure}

In order to better visualize the spatial distribution of the oxygen abundance, and in particular in relation with the locations of the UV-emitting knots, we show in Fig.~\ref{galexn} and \ref{galexs}   the positions of the \hii\/ regions in the northern and southern fields observed by us on top of the GALEX image of M83\footnote{Available at http://photojournal.jpl.nasa.gov/catalog/PIA10374}. We labeled each \hii\/ region (represented by squares) with the \oh\/ value obtained from the \nii/\oii\/ ratio and the \citet{Bresolin:2007} calibration, as well as from the N2 index and the \citet{Pettini:2004} calibration (values in brackets). In Fig.~\ref{galexn} the sudden drop in oxygen abundance that occurs beyond the optical isophotal radius (the dashed curve) is impressively displayed. The six \hii\/ regions at the inner edge of the optical disk have an average \ohtwo8.59 (8.75 if N2 is used), with a very small dispersion, while just beyond this radius the value drops by $\sim$0.30-0.35 dex, and remains virtually constant along the filament that reaches the bright cluster of UV knots at the top of the image.
Qualitatively the same picture is obtained if an alternative metallicity diagnostic, such as \rtwothree, is used.
It is also worth pointing out that the two groups of \hii\/ regions on opposite sides of the \rtf\/ boundary line have similar H$\beta$ luminosities and \oiii/\oii\/ ratios, which implies that the drop in O/H across the optical edge cannot be attributed to systematic differences in ionizing flux output or nebular excitation.

In the southern field (Fig.~\ref{galexs}) the situation could be more complex, depending on the choice of empirical abundances. While also in this case the inner \hii\/ regions near the edge of the bright optical disk have a mean
abundance that matches very well that obtained at the opposite side of the disk [\ohtwo8.62 using \nii/\oii, notice the small variation from this mean for the 10 objects in Fig.~\ref{galexs}], outside \rtf\/ there appears to be a number of nebulae at an intermediate oxygen abundance between that of the inner disk and the outer disk, with \ohthree8.4. These belong to the group of points in Fig.~\ref{abundances} that lie along the exponential fit to the inner gradient. Within the same large clustering of UV knots (left of the bright foreground star) some \hii\/ regions have oxygen abundances that are 0.2 dex smaller, matching those found at the end of the filament near the bottom of the image.
The N2 abundances (in brackets in Fig.~\ref{galexs}) provide, however, a more uniform set of abundances beyond \rtf.
A possibile explanation for the observed variations can be the presence of aperture effects affecting some of the objects, since for this field
the red and blue spectra were obtained on separate nights.

\subsection{Nitrogen}

\begin{figure}
\medskip
\center \includegraphics[width=0.475\textwidth]{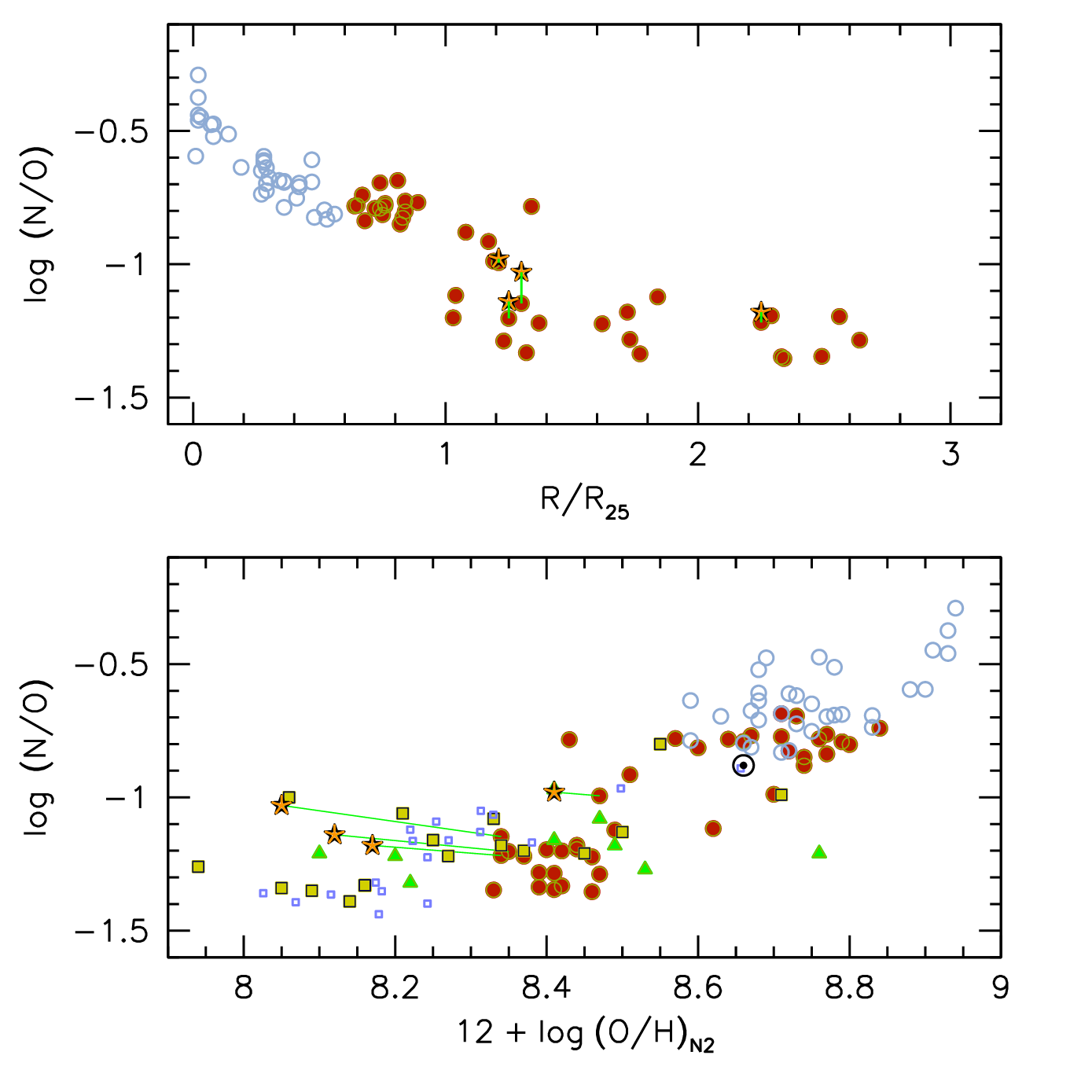}\medskip
\caption{{\it (Top)} Galactocentric gradient of the N/O ratio, expressed in terms of the isophotal radius \rtf. Symbols as in Fig.~\ref{abundances}. Lines connect empirical abundances with \te-based abundances. {\it (Bottom)} Relationship between the N/O ratio and the O/H ratio. The latter has been determined via the N2 diagnostic. A comparison sample has been drawn from \citet[M101, squares]{Kennicutt:2003} and \citet[triangles]{Garnett:1997}. The solar symbol is drawn at 
\oh$_\odot$\,=\,8.66 and log(N/O)$_\odot$\,=\,$-0.88$ (\citealt{Asplund:2005}). For M101 we plot the original \te-based abundances from \citet[full squares]{Kennicutt:2003} and those derived from N2 (for O/H) and Eq.~4 (for N/O, small open squares).\\ \\
 \label{no}}
\end{figure}

In Fig.~\ref{no} we show the radial trend of the N/O ratio ({\it top}) and the N/O ratio as a function of O/H ({\it bottom}). N/O has been measured following the algorithm proposed by \citet{Thurston:1996}, which provides 
the electron temperature in the \nii-emitting zone from the observed R$_{23}$ value:
\\

%\begin{centering}
\begin{equation}
t_2 = 0.6065 + 0.1600x + 0.1878x^2 + 0.2803x^3
\end{equation}
\\[2mm]
%\end{centering}

\noindent
($x=\log$R$_{23}$, $t_2$ in units of 10$^4$~K), and equation (9) by \citet{Pagel:1992} used to derive N$^+$/O$^+$ from the observed \nii/\oii\/ line ratio:
\\

%\begin{centering}
\begin{equation}
\small\log\frac{\rm N^+}{\rm O^+}=\log\frac{\rm [NII]\lambda\lambda6548, 6583}{\rm [OII]\lambda3727} + 0.307 - 0.02\log t_2 - \frac{0.726}{t_2}
\end{equation}
\\[2mm]
%\end{centering}

\noindent
We then assume N$^+$/O$^+$\,=\,N/O. As the top panel of Fig.~\ref{no} shows, the N/O ratio is virtually flat around N/O\,$\simeq$\,$-1.2$ ($\simeq$\,0.5$\times$ the solar ratio) outwards of the isophotal radius. From the bottom panel of Fig.~\ref{no}, where the O/H abundance is obtained from the N2 diagnostic (to avoid using the \nii/\oii\/ ratio for both the
O/H and N/O abundance ratios), we infer that the N/O ratio as a function of O/H throughout the disk of M83 is consistent with the observations in other nearby galaxies, such as the well-studied M101 (squares, corresponding to the \te-based abundances of \citealt{Kennicutt:2003}) and NGC~2403 (\citealt{Garnett:1997}, green triangles). The N/O abundance ratios based on the \oiii\lin4363 auroral-line (open star symbols) are in good agreement, within the scatter, with those determined from R$_{23}$ (also in the case of the M101 \hii\/ regions the agreement is on the order of 0.1 dex). The N/O ratio is approximately constant below \ohtwo8.4, then it rises linearly with O/H. Qualitatively, this is explained in terms of a primary production of nitrogen, which gives rise to the constant pedestal in N/O at low oxygen abundance, and of a secondary component, which is proportional to the oxygen abundance (e.g.~\citealt{Vila-Costas:1993}). We note that the oxygen abundance scale obviously depends on the choice of metallicity diagnostic.

\subsection{Comparison with previous work}\label{comparison}
The work by G07 was the first to characterize the chemical abundances in the XUV disk of M83 (as well as NGC~4625). These authors reported line emission fluxes for 22 \hii\/ regions in this galaxy, 19 of which lie in the outer disk. The chemical analysis was based on a subset of 12 objects (9 in the outer disk), for which reliable fluxes in both the \oii\/ and \oiii\/ lines could be obtained, therefore allowing the determination of the \rtwothree\/ parameter. Lacking \oiii\lin4363 detections, G07 estimated the oxygen abundances from the \rtwothree\/ method, using the calibrations of \citet{McGaugh:1991} and \citet{Pilyugin:2005a}. Apart form the three innermost \hii\/ regions, G07 provide both upper and lower branch abundances for all objects, since, according to these authors, the \rtwothree\/ degeneracy was difficult to remove. While the \nii/\oii\/ and the \nii/\sii\/ line ratios would suggest abundances between \zsun/2 and \zsun\/ (consistent with upper branch abundances), their preferred abundances are between \zsun/10 and \zsun/4 (lower branch), based on photoionization modeling in which the N/O ratio is constant as a function of O/H and equal to the solar value (models with an O/H-scaled value for the N/O ratio yielded less satisfactory fits to the emission
line ratios). This produces a discrepancy in relation to our work: instead of measuring a mean abundance in the outer disk of \oh\,$\approx$\,8.0 (\zsun/5; or lower if using the \citealt{Pilyugin:2005a} method), we measure abundances 
that are between 0.2 dex and 0.6 dex higher (\zsun/3$-$\zsun), depending on the metallicity indicator used.
As a cautionary note, we point out that the O/H abundance ratios derived from N2, O3N2 and \nii/\oii\/
are calibrated using \hii\/  region data and models with a `standard' dependence of N/O on O/H (\citealt{Vila-Costas:1993}). A higher N/O ratio in the outer disk, as proposed by G07, would result in smaller O/H values. 
Our preferred abundances in the outer disk are those that are tied to the \oiii\lin4363 results, i.e.~\oh\,$\approx$\,8.2. 
Our results on the N/O ratio (also those based on the \oiii\lin4363 detections) do not substantiate the choice
made by G07 of a solar N/O ratio in the outer disk of M83: the derived N/O ratios appear normal for the given O/H ratio, given the scatter observed in extragalactic \hii\/ regions (the \te-based N/O ratios appear slightly higher than the average).
Our results are largely independent of the \rtwothree\/ degeneracy, since we have used metallicity indicators that are monotonic with oxygen abundance. 

% - - - - - - - - - - - - - - - - - - - - - - - - - - - -
\section{Discussion}

The main findings  of our observations of the extended disk of M83 can be summarized as follows: {\it (a)} a flat oxygen abundance gradient beyond \rtf, with a relatively high O/H ratio; and {\it (b)} a  discontinuity in the oxygen abundance radial distribution  at the edge of the optical disk. The latter
is most obvious when the oxygen abundance is derived from the \nii/\oii\/ ratio. The discontinuity takes place where the radial abundance gradient turns from exponential to flat. 
This radical change in the abundance distribution just beyond the optical radius of M83 
implies different star formation histories between the inner and outer parts of the disk of this galaxy.

These findings are supported by new emission line data for nearly 50~\hii\/ regions, distributed in galactocentric distance between 0.6 and 2.6 isophotal radii, and by the application of different metallicity diagnostics, including the \oiii\lin4363 auroral line and several strong-line methods.
In this Section we first look at previous detections of changes in the slope of the abundance gradients (what we otherwise call {\it breaks}) in other galaxies. We then critically analyze our abundance data, in order to test whether our results depend on the
choice of metallicity diagnostics. Finally, we attempt to provide an interpretation for the observed features in the abundance distribution.

\subsection{Bimodal abundance gradients: observations and theory}
Changes in galactic abundance gradients, with shallower (or flat) slopes at large galactocentric distances, have already been determined from studies of the \hii\/ regions located in the main optical disk of a few spiral galaxies. For example, \citet{Martin:1995} found a break in the slope of the abundance gradient of the barred galaxy NGC~3359, in correspondence of the corotation radius. A similar break was observed by \citet{Roy:1997} in NGC~1365, also a barred galaxy. In both these cases, the break in the gradient occurs well within the bright optical disk of the galaxy. In M83, instead, we find that the flattening occurs just outside of the main star-forming disk, approximately at the isophotal radius.

A bimodal chemical abundance distribution is observed also in the Milky Way. The Cepheid abundances of various elements, and iron in particular, show a discontinuity (on the order of 0.2 dex) in their radial distribution, with a shallower gradient at galactocentric distances larger than 10~kpc (\citealt{Luck:2003,Andrievsky:2004,Yong:2006}). A similar discontinuity has been observed by \citet{Twarog:1997} in the metallicity distribution of Galactic open clusters. The lack of efficient radial mixing beyond the corotation resonance (where the spiral pattern speed equals the galactic rotation velocity) and the gap in gas density at corotation could be responsible for the break in the abundance gradient and for the discontinuity determined from the Cepheid observations (\citealt{Mishurov:2002,Lepine:2003,Acharova:2005}). 
The behavior of the ionized gas abundances is less clear, since \citet{Vilchez:1996} obtained an essentially flat gradient in the nitrogen and oxygen abundances beyond 12~kpc from the center, and \citet{Maciel:1999} measured a similar flattening for oxygen and other elements from planetary nebulae located beyond 10~kpc, 
while other authors (e.g.~\citealt{Deharveng:2000,Rudolph:2006}) found no evidence for breaks 
in the \hii\/ region abundance gradient.

Turning to the theoretical side, non-linear gradients (in logarithmic abundances), flatter at large galactocentric distances, can be reproduced by `inside-out' scenarios of galaxy formation in which the Milky Way disk is built up via gas infall, and in which the timescale for the formation of the disk increases with galactocentric distance (\citealt{Matteucci:1989,Boissier:1999,Chiappini:2003}). In these models the radial variations of the gas infall rate and of the star formation rate in the galactic disk control the formation and the slope of the abundance gradients. As shown by \citet{Chiappini:2001}, the shape of the abundance gradients at large galactocentric distances is regulated by the evolution of the halo.
Despite the fact that different authors predict that the overall disk abundance gradients become either flatter (e.g.~\citealt{Prantzos:2000} and \citealt{Hou:2000}, who also predict a flatter gradient in the {\it inner} disk) or steeper (e.g.~\citealt{Chiappini:2001}) with time, recent galactic evolution models are able to provide a good match to the radial trends of many different chemical species measured for Cepheids, young stars and open clusters (\citealt{Cescutti:2007}). The development of a plateau in the abundance gradient at large galactocentric distance (as well as in the central region) is also predicted by the chemodynamical models of \citet{Samland:1997}. For galaxies beyond the Milky Way, however, it is hard to generalize these model results, since flat abundance gradients and/or discontinuities are rarely, if ever, observed. In fact, almost no information on nebular abundances beyond the optical edges of spiral disks is available.
In summary, changes in the abundance gradients, with flatter slopes in the outer regions, as well as abundance discontinuities, have been observed in other galaxies, although very rarely (a discontinuity only in the Milky Way). Despite some qualitative similarities with the examples mentioned above,  we cannot conclude that the physical causes of the abundance distribution in M83 are necessarily the same.

\subsection{Is the abundance break in M83 real?}
\begin{figure}
\medskip
\center \includegraphics[width=0.475\textwidth]{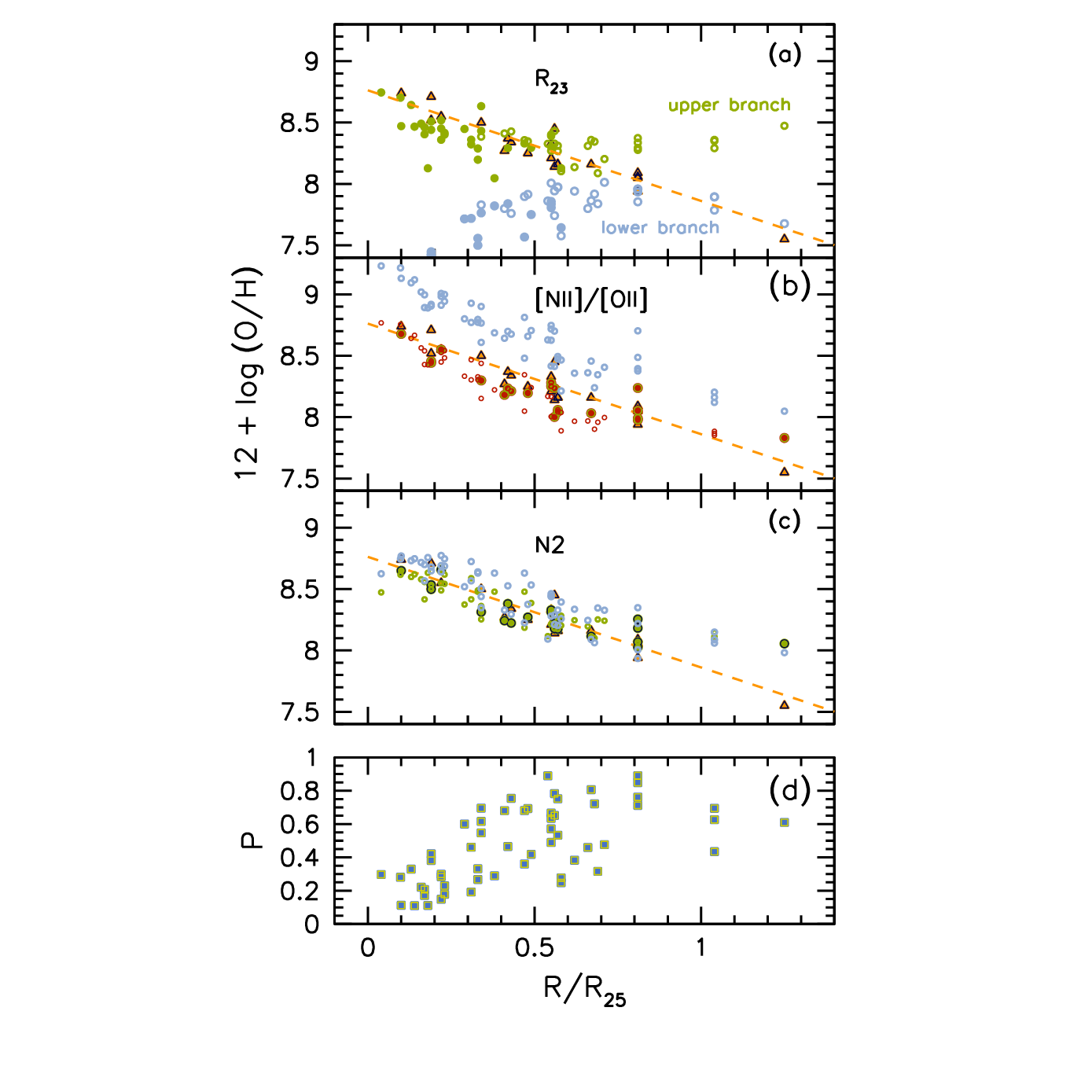}\medskip
\caption{Abundance gradient in M101 from different strong-line indicators: {\it (a)} R$_{23}$, showing the results of the upper branch (green) and lower branch (light blue) calibrations of \citet{Pilyugin:2005a}. Full dots correspond to values of the N2 indicator larger than $-1.0$. 
{\it (b)} The \nii/\oii\/ indicator, calibrated by \citet[light blue circles]{Kewley:2002} and \citet[red dots]{Bresolin:2007}. Larger symbols represent \hii\/ regions with available \oiii\lin4363 auroral line measurements.
{\it (c)} N2, calibrated by \citet[light blue circles]{Denicolo:2002} and \citet[green dots]{Pettini:2004}. 
Larger symbols represent \hii\/ regions with available \oiii\lin4363 auroral line measurements.
The triangles represent \hii\/ regions with oxygen abundances obtained from the \lin4363 line. The dashed line is the abundance gradient obtained from a linear fit to these
abundances.
{\it (d)} The radial trend of the P parameter in M101. The observational data are from \citet{Kennicutt:2003}, \citet{Bresolin:2007}, \citet{van-Zee:1998a} and \citet{Kennicutt:1996}.\\
\label{m101}}
\end{figure}

\citet{Pilyugin:2003c} has questioned the reality of slope changes in the radial abundance gradients of spiral galaxies, showing that an artificial break can occur when abundances are determined via the R$_{23}$ method, either as a result of adopting the wrong branch of the indicator outside of a certain galactocentric distance, or as due to a break in the radial trend of the excitation parameter P\,=\,\oiii/(\oii+\oiii). We think that our result in the case of M83 is quite robust, having inferred the abundance break from a variety of metallicity indicators, most of which are monotonic with O/H, thus being unaffected by the degeneracy of the R$_{23}$ method. To strengthen our conclusion, we proceeded with a test similar to that carried out by \citet{Pilyugin:2003c}, i.e.~we applied our abundance indicators to the \hii\/ regions of the galaxy M101, for which \citet{Pilyugin:2003c} showed that the use of the R$_{23}$ indicator can lead to a fictitious break in the gradient. The results of our test are shown in Fig.~\ref{m101}. In panel {\it (a)} the oxygen abundance has been obtained from the R$_{23}$ diagnostic and the P-method, as calibrated by \citet{Pilyugin:2005a}. The green symbols refer to the upper branch calibration, while the light blue symbols refer to the lower branch calibration. Full dots are used for N2\,$>$\,$-1$, open circles 
for N2\,$<$\,$-1$. This plot shows that, indeed, the adoption of a single, upper branch calibration of R$_{23}$ throughout the whole disk produces a radial trend that deviates in the outer galactic region from the linear gradient measured from the \oiii\lin4363 auroral line (triangles and dashed line, from \citealt{Kennicutt:2003}), here chosen to represent the real radial trend of the oxygen abundance. The break occurs approximately 
at the radius in which a break in the P parameter occurs [panel {\it (d)}].
The N2-based criterion to discriminate between the upper and lower branches appears effective at the extremes of the abundance range, while in the intermediate range 8.0\,$<$\,\oh\,$<$\,8.3 the branch selection from N2 is less in agreement with the \lin4363-based results. This is not unexpected, since the P-method is not calibrated within this range (\citealt{Pilyugin:2005a}). 

The abundances in panel {\it (b)} of Fig.~\ref{m101} were obtained from the \nii/\oii\/ ratio, as calibrated by \citet[light blue circles]{Kewley:2002} and \citet[red dots]{Bresolin:2007}. In the latter case, we used the larger symbols for \hii\/ regions with available \oiii\lin4363 abundances (\citealt{Kennicutt:2003}). Apart from  the systematic offset noted previously, the two calibrations provide virtually the same abundance gradient, with a slope that is in good agreement with the \lin4363-based one, and without any obvious deviation from a linear trend.
A possible decrease in the gradient slope of the strong-line abundances, suggested by the single \hii\/ region 
at $R$/\rtf\,=\,1.25, might occur at \oh\,$<$\,8.0 in the \citet{Bresolin:2007} scale, i.e.~well below the abundance at which the gradient in M83 levels off to a constant O/H. This suggests that the relative abundances determined in M83 from the \nii/\oii\/ ratio are reliable throughout the inner and the outer disk.
The same conclusion can be reached from panel {\it (c)} of Fig.~\ref{m101}, where the N2 indicator was used.
We show the abundances from the calibrations of \citet[green dots]{Pettini:2004} and \citet[open light blue circles]{Denicolo:2002}. Once again, the agreement with the abundances from the auroral-lines is good down to \ohtwo8.2, while the N2 abundances in the outer disk of M83 flatten around \ohtwo8.4 (see Fig.~\ref{abundances}).
In conclusion, this test supports our results concerning the abundance gradient in M83, and in particular the
fact that the slope change in the abundance gradient seen at the isophotal radius is real.

\subsection{Explaining the abundance distribution in M83}
If the abundance gradient in M83 has the shape shown in Fig.~\ref{abundances}, with both a flat trend and a  discontinuity occurring around the isophotal radius, it is important to understand how these features can be explained within a model of galactic evolution. The fact that nowhere in the extended disk of the galaxy the oxygen abundance is below 1/3 solar (or possibly even higher, depending on the abundance diagnostic that is used) is also remarkable. Because of the low levels of star formation activity at large galactocentric distances, as measured from the averaged \halpha\/ profile, we would expect low rates of chemical enrichment and therefore rather low oxygen abundances in the outskirts of the galaxy. Our findings go against this simple expectation, indicating that some processes are at work that both homogenize and increase the chemical abundances even far from the region of main galactic star-forming activity. However, a non-negligible level of chemical enrichment in extended spiral disks is expected, based on the observed UV emission (which measures star formation on timescales on the order of a few hundred Myr), as well as on the presence of older generations of stars that are resolved beyond the optical edges of nearby late-type spirals (e.g.~NGC~2403, NGC~300, M33: \citealt{Davidge:2003,Bland-Hawthorn:2005,Barker:2007}). 
Moreover, the discovery of intergalactic \hii\/ regions (\citealt{Gerhard:2002,Ryan-Weber:2004}) has led to the determination of near-solar gas-phase oxygen abundances at projected distances up to 25-30~kpc from the nearest galaxies (\citealt{Mendes-de-Oliveira:2004}). In at least some of these cases the star forming activity appears to have been triggered by tidal interactions between galaxies, which could also help explain the high metallicity as a result of chemically enriched gas being stripped out of metal-rich galaxy disks  (\citealt{Sakai:2002}).
A crude estimate of the expected present-day gas abundances in the outer disk of M83  can be 
obtained by assuming that star formation has been ongoing for at least the last Gyr, as deduced 
by \citet{Dong:2008}. However, numerical simulations by \citet{Bush:2008} suggest that
star formation  is a long-lasting process in XUV disks, extending over timescales that can be as long as the lifetime of the galaxy.
We have estimated the current SFR density ($\Sigma_{SFR}$) of the outer disk of M83 by measuring 
the far-UV  flux detected in the form of UV clusters and complexes  beyond \rtf\/ and out to 4\,\rtf. The value we found, after correcting for the presence of older clusters and background sources, is $\Sigma_{SFR}=1.0\times10^{-5}$\,\msun\,yr$^{-1}$\,kpc$^{-2}$, 
much lower than the estimate of $8\times10^{-4}$\,\msun\,yr$^{-1}$\,kpc$^{-2}$ obtained by \citet{Dong:2008} 
from the 8\,$\mu$m emission in two Spitzer IRAC fields at approximately 2.3\,\rtf\/ from the center. We attribute much of the discrepancy to the fact that we consider the SFR integrated over the whole outer disk (out to 4\,\rtf), therefore averaging the far-UV emission from star-forming complexes with the null contribution from large zones where no clusters are detected. The two fields in \citet{Dong:2008} were selected based on the presence of active star forming regions, and have therefore an
enhanced $\Sigma_{SFR}$ relative to the mean of the outer disk. Our lower value is thus more representative of the SFR of the outer disk as a whole, and would be more appropriate to adopt when considering an evolutionary scenario in which sites of star formation appear stochastically throughout the
extended disk, perhaps as a consequence of triggering by outer spiral perturbations  (\citealt{Bush:2008}).

With the current rate of star formation, as determined above, and a net oxygen yield of 0.01 (\citealt{Maeder:1992}),  
$\sim$\,$10^2$\,\msun\/ of oxygen would have been released per kpc$^2$ into the interstellar medium of the outer disk over the past Gyr. With a gas surface density $\Sigma$(\hi)\,$\simeq$\,1.6\,\msun\,pc$^{-2}$ at 3\,\rtf\/ (derived combining information from \citealt{Rogstad:1974}, \citealt{Huchtmeier:1981} and \citealt{Crosthwaite:2002}), we then estimate an O/H ratio roughly equal to 1/7 of the solar value [\ohtwo7.8] as a result of the chemical enrichment over the past Gyr.
Considering the approximations and uncertainties involved in this estimate, we can therefore reasonably recover the oxygen abundance measured in the extended disk of M83 by assuming a constant SFR over the past 2-3 Gyr. Longer timescales, up to several Gyr, can be easily accomodated, for example by assuming lower SFR in the past than the current value, or smaller oxygen yields (as in ~\citealt{Pilyugin:2007a}).
This demonstrates that a low, but persistent, level of star formation can lead to a substantial chemical enrichment, compatible with the observed abundances. 
We stress that this association between gas density, star formation and metallicity can only be made in regions where star formation has been taking place. There may well be areas of the outer disk that have sub-critical gas densities and are therefore quiescent and metal poor  if they have remained below the critical density during
most of their history.

We also point out that, considering the abundances of the outer disk of M83 within the context of a closed box chemical evolution model, the effective yield $y_{\mbox{\footnotesize eff}}=Z/(ln\,\mu^{-1})$ (\citealt{Edmunds:1990})
would be quite large in the extended disk of M83, considering that the gas mass fraction $\mu$ must be near unity (although no estimate of the stellar mass is available, its fractional component is presumably very small, given the absence of an easily detectable stellar disk). In other words, the outskirts of M83 are overabundant for their gas fraction, the opposite of what is observed in dwarf galaxies (e.g.~\citealt{Matteucci:1983,van-Zee:2006a}).\\

As noted above, gas-phase abundance plateaus have been inferred for a few barred galaxies from the measurement of strong-line indexes (\citealt{Martin:1995,Roy:1997}). In these cases the explanation of the observed breaks in abundance gradients is given in terms of the non-axisymmetric potential of a young and strong bar, in which radial gas flows produce, with time, increasingly shallower slopes across the galactic disk, as a consequence of the redistribution of mass and angular momentum, and a `steep-shallow' transition 
at the galactocentric distance that corresponds approximately to the corotation radius (\citealt{Friedli:1994,Friedli:1998}). 
Gas flows that are preferentially directed outwards beyond corotation have a diluting effect, producing a flattened abundance distribution in the outer part of the disk.  

The bar in M83, with a deprojected axis ratio $(b/a)=0.38$ (\citealt{Martin:1995a}), is among the strongest (lowest $b/a$) in the sample of barred galaxies studied by \citet{Martinet:1997}, who show that stronger bars correlate with shallower radial abundance gradients. According to \citet{Martin:1995}, the weak star formation in the bar, the strong circumnuclear activity and the shallow abundance gradient, as observed in the disk of M83, support the idea 
that this is a rather evolved barred system. Within a picture of secular evolution of barred galaxies, an initial phase of intense star formation, that produces a discontinuity in the abundance gradient at the edge of the bar, is followed by an almost quiescent phase in the bar, during which the funneling of gas towards the central regions triggers star formation near the nucleus (\citealt{Combes:2008}). Recent CO (\citealt{Lundgren:2004}) and \halpha\/ (\citealt{Fathi:2008}) kinematical data have confirmed the presence of a gas inflow along spiral trajectories, from kpc-size scales down to a few tens of pc from the nucleus of M83, and which is likely driven by the potential of the bar.

While the gas flows induced by the presence of the bar are known to generate overall flatter abundance gradients in the inner disk of 
spiral galaxies compared to non-barred galaxies (\citealt{Vila-Costas:1992, Zaritsky:1994, Dutil:1999}), this mechanism seems unlikely, or at least insufficient, to explain the abundance trend in the outer disk of M83. As mentioned above, the abundance break caused by a bar is predicted to occur at the 
galactocentric distance that corresponds to corotation.
In the case of M83, corotation is at 2.83 arcmin from the center (3.7 kpc, \citealt{Lundgren:2004}), while the break in abundance gradient occurs much further out, at the edge of the optical disk (\rtf\,=\,6.44 arcmin, 8.4~kpc). 
On the other hand, angular momentum redistribution and gas flows predicted for viscous evolution of star-forming disks (\citealt{Lin:1987, Clarke:1989, Ferguson:2001})
can give rise to flat gradients in the ISM of the outer parts of spiral disks (\citealt{Sommer-Larsen:1990,Tsujimoto:1995,Thon:1998}).\\

\begin{figure}
\medskip
\center \includegraphics[width=0.475\textwidth]{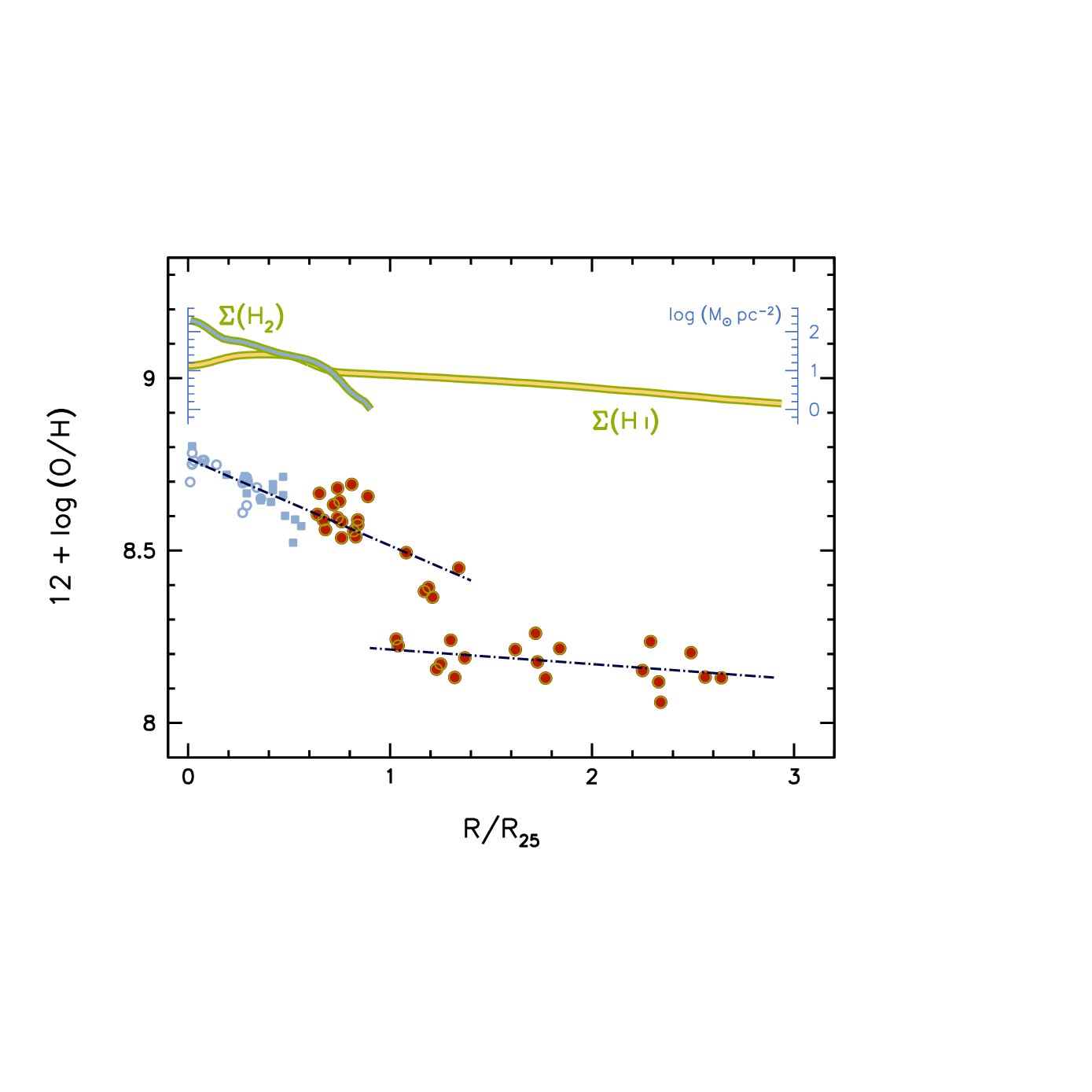}\medskip
\caption{Comparison between the empirical O/H gradient (from Fig.~\ref{abundances}e) and the logarithmic gas density profile (units for the latter are shown at the top right). The two top curves represent the surface density distribution of \hi\/ (from the model of \citealt{Rogstad:1974}) and H$_2$ (from \citealt{Crosthwaite:2002}). \\
\label{hi}}
\end{figure}

An alternative explanation can be explored, by noting that in M83 the drop in the azimuthally averaged \halpha\/ surface brightness at \rhii\/ occurs in proximity  of the outer Lindblad resonance, at 4.83 arcmin (6.3~kpc, \citealt{Lundgren:2004}), which could act as a possible potential barrier for gas flows, and which might cause discontinuities in the star formation rate (\citealt{Andrievsky:2004}). 
We also note that the abundance trends shown in Fig.~\ref{abundances} mimic the total gas (neutral\,+\,molecular) surface density profile shown by \citet[their Fig.~8]{Crosthwaite:2002}, which flattens at approximately 6 arcmin from the center, beyond the sharp decline in both CO and \hi\/ emission at the edge of the optical disk  (but we must point out that molecular data have not yet been published for the disk outside \rtf). \citet{Crosthwaite:2002} pointed out that the abrupt discontinuity between the molecular-dominated inner disk ($r<5'$) and the atomic outer envelope is rather unique among late-type spiral galaxies. In relation to this finding, a modest drop in the mean hydrogen volume densities of giant molecular clouds at $R>$\,\rtf\/  has recently been measured by \citet{Heiner:2008}. Therefore, the flat abundance gradient we observe in the outskirts of M83 may simply be a reflection of the gas density profile beyond the optical radius. This is illustrated in Fig.~\ref{hi}, where we show the empirical abundances taken from Fig.~\ref{abundances}(e), and the gas density profiles from \citet[their model \hi\/ distribution in Fig.~7]{Rogstad:1974} and \citet[their H$_2$ data]{Crosthwaite:2002}. As the comparison shows, the relative slopes of the inner and outer O/H distributions closely match the relative slopes of the $\Sigma$(H$_2$) and $\Sigma$(\hi) profiles, respectively. The 
discontinuity in the O/H distribution occurs where the \hi\/ surface density starts to dominate the radial gas profile.

Beyond the optical edge, a large envelope of neutral hydrogen is present (\citealt{Rogstad:1974,Huchtmeier:1981}), extending  47~arcmin (62~kpc) from the center, and containing an \hi\/ ring outside the optical radius (\citealt{Tilanus:1993}). The low gas surface density in the envelope agrees with the low level of star formation in the outer disk. Under such a situation, with low efficiencies of the processes that lead to star formation, 
flat abundance gradients are expected, due to the unevolved status of the system (as shown, or example, by the models of \citealt{Molla:2005}). The situation might be similar to that found in low surface brightness galaxies (LSBGs). A possible analogy between XUV disks and the gas-rich LSBGs has already been noted by \citet{Thilker:2007}. The star forming activity in LSBGs takes place at low, possibly subcritical, gas densities (\citealt{van-der-Hulst:1993,de-Blok:1996}). Low rates of star formation, perhaps occuring sporadically or intermittently, have been obtained for LSBGs by several authors (\citealt{Gerritsen:1999, van-den-Hoek:2000, Boissier:2008}). The low chemical abundances of LSBGs (\citealt{McGaugh:1994}) denote fairly unevolved systems, even if there are exceptions (\citealt{Bergmann:2003}). The notion of LSBGs as `slowly evolving' systems (\citealt{McGaugh:1997}) might then fit the description of outer spiral disks, as well. The \hii\/ region oxygen abundances measured in LSBGs by \citet{McGaugh:1994}, \citet{de-Blok:1998} and \citet{Kuzio-de-Naray:2004} span a wide range, \ohtwo$7.6-8.8$, without a preferred value (these O/H abundances, obtained from \rtwothree, are in the \citealt{McGaugh:1991} theoretical scale). The oxygen abundance value obtained in the extended disk of M83 lies comfortably within this range and, as shown earlier, is consistent with the enrichment level one obtains with the present SFR in the outer disk 
over timescales of a few Gyr.
Moreover, the work by \citet{de-Blok:1998} showed that the radial distribution within LSBGs galaxies is approximately flat, i.e. there is no evidence for the presence of gradients in the oxygen abundance. 
As suggested by these authors, this finding would agree with the sporadic star formation rates measured in LSBGs. 
Star formation in low surface brightness disk galaxies, as well as in the outer disk of M83, seems to favor a chemical evolution in which large-scale gradients are not produced. 

The wide variety of metal enrichment seen in LSBGs, as well as dwarf galaxies, may reflect their range of morphologies: some have a quite obvious extended disk structure (e.g.~NGC6822, \citealt{Weldrake:2003}), while others are much more irregular. Perhaps it is the LSBGs with rotationally supported disks that can maintain slow, but steady star formation. In the case of the outer disk of M83, \citet{Dong:2008} find that 
the \hi\/ surface density distribution peaks at the critical density (but we note again that the total H$_2$\,+\,\hi\/   surface density of the outer disk is not known, due to the lack of molecular data).
\\

\rm
Finally, we should also entertain the idea that peculiarities in the chemical abundance distribution
of the extended disk of M83 could arise from the gravitational interaction with one or more of the several dwarf galaxies found in the vicinity of M83 (\citealt{Karachentsev:2007}). An interaction with
NGC~5253 about 1~Gyr ago has been proposed by \citet[see also \citealt{Calzetti:1999}]{van-den-Bergh:1980}.  
The faint dwarf KK~208 (\citealt{Karachentseva:1998}), 25~kpc in projection to the north of the center of M83,  
is strongly elongated, likely as a result of the tidal interaction with M83 (\citealt{Karachentsev:2002}). Its location does not correspond to any of the UV-bright filaments discovered by GALEX, but UV knots are visible at even larger galactocentric distances.

The warped structure of the extended \hi\/ envelope of M83 (\citealt{Rogstad:1974})
is a possible signature of past galaxy encounters. Moreover, from the kinematic point of view the outer \hi\/ envelope appears to be decoupled from 
the inner part, with disk rotation taking place on different planes (\citealt{Huchtmeier:1981}).
Gas stripping and/or redistribution taking place as an effect of the interactions could help explain the different chemical properties betweeen the inner and the outer 
part of the disk. For instance, we could explain the high
metal abundance of the outer disk if we postulated that it is mainly composed of gas that was tidally transported from chemically evolved regions at smaller radii in the main disk of M83.
\\

% - - - - - - - - - - - - - - - - - - - - - - - - - - - -
\section{Conclusions}

The outskirts of spiral galaxies hold a powerful diagnostic value in the study of their formation and evolution.
Breaks in the exponential surface brightness profiles are detected up to six disk scale lengths (\citealt{Pohlen:2006, Erwin:2008}), reflecting at least in some cases breaks in the mass distribution (\citealt{Bakos:2008}).
These data are spawning a number of models for their interpretation, and include the effect of radial star formation drops resulting from drops in the gas density (\citealt{Roskar:2008}), star formation thresholds (\citealt{Elmegreen:2006}) and angular momentum redistribution by galactic bars (\citealt{Debattista:2006}).
In the UV-detected extended spiral disks star formation takes place even beyond these breaks in the light profiles.
The chemical abundance analysis of these outer disk regions therefore provides essential information on the processes that lead to the formation of galactic disks.

We have studied the oxygen abundance gradient in the outer disk of M83, in the radial span between 0.6 and 2.6 times the isophotal radius, obtaining deep optical spectra for nearly 50 \hii\/ regions. We have obtained nebular O/H ratios adopting different strong line metallicity indicators, as well as the direct method that involved the use of the \oiii\lin4363 auroral line.
Regardless of the abundance indicator used, we obtain a flat oxygen abundance gradient beyond the \rtf\/ isophotal radius, and a drop in abundance slightly beyond this galactocentric distance. The value of the approximately constant abundance in the outer disk depends on the adopted diagnostic, and 
varies between \ohtwo8.2 and \ohtwo8.6.

By analogy with the results obtained for LSBGs, we can interpret the observed abundances in the outer disk of M83
as due to its relatively unevolved state. The slow evolution of the outer disk, perhaps still in the process of being assembled via gas inflow, results in a flat distribution of chemical abundances outside the region of main star formation. The rather abrupt change in star formation properties and gas density between the inner and 
outer disk produces the observed break near the optical edge of the galaxy.

Generalizing the results obtained for M83 is, at this point, obviously premature. 
A conclusion on whether the trends observed in M83 are a feature of the majority of extended spiral disks, or are instead peculiar to this galaxy, awaits further detailed chemical abundance studies in a larger number of similar systems.

% - - - - - - - - - - - - - - - - - - - - - - - - - - - -
\acknowledgments
FB would like to thank the Institute of Astronomy,  and Elena and Roberto Terlevich in particular, for the warm hospitality in Cambridge while most of this work was carried out, and gratefully acknowledges partial support for this work from the National Science Foundation grant 0707911. We also wish 
to thank Marina Rejkuba and Magda Arnaboldi at the ESO User Support Group for their assistance with our observations.
We are grateful to an anonymous referee for his/her careful reading of the manuscript.
This research has made use of NASA's Astrophysics Data System Bibliographic Services.\\

\noindent
{\it Facilities:} \facility{VLT:Antu (FORS2)}\\
{\it Facilities:} \facility{Bok (90Prime)}

%\clearpage

\appendix{}
We summarize in Table~\ref{galaxies} the equatorial positions (Columns 2 and 3) and the redshifts (Column 4) of background emission-line galaxies that were serendipitously discovered in this study. The numbering scheme for the FORS field in Column~5  is the same as in Fig.~\ref{image}.
\input{tab4}

\bibliography{M83}

\end{document}

%% file: tab1.tex
\begin{deluxetable}{ccccc}
\tabletypesize{\scriptsize}
\tablecolumns{5}
%\tablewidth{300pt}
\tablewidth{0pt}
\tablecaption{H\,\scriptsize II \small region sample\label{sample}}

\tablehead{
\colhead{\phantom{aaaaa}ID\phantom{aaaaa}}	     &
\colhead{R.A.}	 &
\colhead{Decl.}  &
\colhead{R/R$_{25}$}	 &
\colhead{Notes}	 \\[0.5mm]
\colhead{}       &
\colhead{(J2000.0)}   &
\colhead{(J2000.0)}   &
\colhead{}   &
\colhead{} \\[1mm]
\colhead{(1)}	&
\colhead{(2)}	&
\colhead{(3)}	&
\colhead{(4)}	&
\colhead{(5)}	}
\startdata
\\[-1mm]
1\dotfill  & 13~ 35~ 56.71  &  $-$29~ 56~ 37.3  &   2.33 &	      \\       %  2_01
2\dotfill  & 13~ 36~ 12.45  &  $-$29~ 57~ 06.6  &   1.84 &	      \\       %  2_18
3\dotfill  & 13~ 36~ 13.27  &  $-$29~ 52~ 44.6  &   1.68 &	      \\       %  2_11
4\dotfill  & 13~ 36~ 19.25  &  $-$29~ 53~ 43.8  &   1.47 &	      \\       %  2_19
5\dotfill  & 13~ 36~ 28.58  &  $-$29~ 56~ 06.0  &   1.28 &	      \\       %  2_31
6\dotfill  & 13~ 36~ 29.23  &  $-$29~ 54~ 12.2  &   1.15 &	      \\       %  2_28
7\dotfill  & 13~ 36~ 29.30  &  $-$29~ 54~ 34.1  &   1.16 &	      \\       %  2_29
8\dotfill  & 13~ 36~ 31.08  &  $-$29~ 55~ 42.0  &   1.17 &	      \\       %  2_33
9\dotfill  & 13~ 36~ 32.53  &  $-$29~ 54~ 13.0  &   1.04 &	      \\       %  2_32
10\dotfill & 13~ 36~ 46.55  &  $-$29~ 46~ 40.5  &   1.03 &	      \\       %  1_24
11\dotfill & 13~ 36~ 46.75  &  $-$29~ 44~ 42.0  &   1.32 &	      \\       %  1_17
12\dotfill & 13~ 36~ 47.71  &  $-$29~ 46~ 28.1  &   1.04 &	      \\       %  1_23
13\dotfill & 13~ 36~ 49.08  &  $-$29~ 56~ 02.9  &   0.76 &	      \\       %  3_35
14\dotfill & 13~ 36~ 50.07  &  $-$29~ 55~ 52.7  &   0.72 &	      \\       %  3_34
15\dotfill & 13~ 36~ 50.25  &  $-$29~ 42~ 28.5  &   1.62 &	      \\       %  1_09
16\dotfill & 13~ 36~ 51.10  &  $-$29~ 56~ 13.1  &   0.75 &	      \\       %  3_33
17\dotfill & 13~ 36~ 53.09  &  $-$29~ 55~ 50.0  &   0.67 &	      \\       %  3_32
18\dotfill & 13~ 36~ 54.24  &  $-$29~ 47~ 13.8  &   0.82 &	      \\       %  1_28
19\dotfill & 13~ 36~ 54.79  &  $-$29~ 47~ 07.2  &   0.83 &	      \\       %  1_27
20\dotfill & 13~ 36~ 55.17  &  $-$30~ 05~ 47.4  &   2.25 &  XUV 01  \\       %  4_19
21\dotfill & 13~ 36~ 55.35  &  $-$29~ 47~ 32.1  &   0.76 &	      \\       %  1_30
22\dotfill & 13~ 36~ 55.64  &  $-$29~ 47~ 37.6  &   0.74 &	      \\       %  1_31
23\dotfill & 13~ 36~ 55.69  &  $-$29~ 55~ 46.4  &   0.64 &	      \\       %  3_30
24\dotfill & 13~ 36~ 56.79  &  $-$29~ 55~ 52.2  &   0.65 &	      \\       %  3_29
25\dotfill & 13~ 36~ 57.23  &  $-$30~ 08~ 12.0  &   2.64 &	      \\       %  4_11
26\dotfill & 13~ 36~ 57.59  &  $-$29~ 41~ 07.7  &   1.77 &	      \\       %  1_01
27\dotfill & 13~ 36~ 58.07  &  $-$30~ 07~ 40.2  &   2.56 &	      \\       %  4_12
28\dotfill & 13~ 36~ 58.17  &  $-$30~ 07~ 17.1  &   2.49 &	      \\       %  4_13
29\dotfill & 13~ 36~ 58.52  &  $-$29~ 59~ 24.5  &   1.21 &  XUV 13  \\       %  3_27
30\dotfill & 13~ 36~ 58.64  &  $-$30~ 06~ 00.5  &   2.29 &  XUV 02  \\       %  4_18
31\dotfill & 13~ 36~ 59.35  &  $-$29~ 41~ 18.7  &   1.73 &	      \\       %  1_02
32\dotfill & 13~ 36~ 59.62  &  $-$30~ 06~ 21.0  &   2.34 &	      \\       %  4_17
33\dotfill & 13~ 37~ 00.29  &  $-$29~ 41~ 24.0  &   1.72 &	      \\       %  1_03
34\dotfill & 13~ 37~ 00.39  &  $-$29~ 47~ 45.9  &   0.68 &	      \\       %  1_32
35\dotfill & 13~ 37~ 00.49  &  $-$29~ 57~ 23.2  &   0.89 &	      \\       %  3_24
36\dotfill & 13~ 37~ 01.19  &  $-$29~ 57~ 10.8  &   0.86 &	      \\       %  3_23
37\dotfill & 13~ 37~ 01.69  &  $-$30~ 00~ 54.3  &   1.55 &  XUV 06  \\       %  3_22
38\dotfill & 13~ 37~ 04.47  &  $-$29~ 46~ 55.9  &   0.81 &	      \\       %  1_26
39\dotfill & 13~ 37~ 04.99  &  $-$29~ 59~ 45.9  &   1.30 &  XUV 09  \\       %  3_19
40\dotfill & 13~ 37~ 05.82  &  $-$30~ 00~ 11.2  &   1.37 &	      \\       %  3_18
41\dotfill & 13~ 37~ 06.76  &  $-$29~ 59~ 57.0  &   1.34 &  XUV 08  \\       %  3_17
42\dotfill & 13~ 37~ 07.13  &  $-$29~ 56~ 47.6  &   0.84 &  XUV 22  \\       %  3_16
43\dotfill & 13~ 37~ 08.20  &  $-$29~ 59~ 19.6  &   1.25 &  XUV 11  \\       %  3_15
44\dotfill & 13~ 37~ 08.93  &  $-$29~ 56~ 39.8  &   0.84 &	      \\       %  3_14
45\dotfill & 13~ 37~ 09.74  &  $-$29~ 59~ 04.5  &   1.23 &	      \\       %  3_13
46\dotfill & 13~ 37~ 10.76  &  $-$29~ 55~ 45.4  &   0.74 &	      \\       %  3_12
47\dotfill & 13~ 37~ 12.36  &  $-$29~ 57~ 52.9  &   1.08 &	      \\       %  3_11
48\dotfill & 13~ 37~ 16.42  &  $-$30~ 00~ 47.6  &   1.59 &	      \\       %  3_06
49\dotfill & 13~ 37~ 19.35  &  $-$29~ 57~ 40.3  &   1.19 &	      \\       %  3_02
\enddata
\tablecomments{Units of right ascension are hours, minutes and seconds, and units of declination
are degrees, arcminutes and arcseconds. Col.~(1): \hii\/ region identification. Col.~(2): Right
Ascension. Col.~(3): Declination. Col.~(4): Galactocentric distance in units of R$_{25}$\,=\,6.44 arcmin.
Col.~(5): Identification from G07.\medskip}
\end{deluxetable}

%% file: tab2.tex
\begin{deluxetable*}{cccccccc}
\tabletypesize{\scriptsize}
\tablecolumns{8}
\tablewidth{0pt}
\tablecaption{Reddening-corrected fluxes\label{fluxes}}

\tablehead{
\colhead{\phantom{aaaaa}ID\phantom{aaaaa}}	     &
\colhead{\oii}	 &
\colhead{\oiii}  &
\colhead{\nii}	 &
\colhead{\sii}	 &
\colhead{\ariii}	 &
\colhead{F(H$\beta$)} &
\colhead{$c$(\hbeta)}   \\[0.5mm]
\colhead{}       &
\colhead{3727}   &
\colhead{5007}   &
\colhead{6583}   &
\colhead{6717+6731} &
\colhead{7135}  &
\colhead{(erg\,s$^{-1}$\,cm$^{-2}$)}     &
\colhead{(mag)}     \\[1mm]
\colhead{(1)}	&
\colhead{(2)}	&
\colhead{(3)}	&
\colhead{(4)}	&
\colhead{(5)}	&
\colhead{(6)}	&
\colhead{(7)}   &
\colhead{(8)}    }
\startdata
\\[-1mm]
 1\dotfill &   370 $\pm$   20 &     146 $\pm$    7 &     36 $\pm$    3 &     28 $\pm$    2 &           \nodata    &  1.7 $\times 10^{-15}$ &  0.00 \\   %2_01
 2\dotfill &   472 $\pm$   59 &     197 $\pm$   13 &     63 $\pm$   12 &     78 $\pm$   15 &    0.0 $\pm$ 15.8    &    3.0 $\times 10^{-16}$ &  0.08 \\   %2_18
 3\dotfill &   \nodata        &      29 $\pm$    4 &     60 $\pm$   12 &     69 $\pm$   13 &    0.0 $\pm$ 11.3    &    1.1 $\times 10^{-16}$ &  0.17 \\   %2_11
 4\dotfill &   \nodata        &     162 $\pm$    7 &     44 $\pm$    3 &     53 $\pm$    3 &    9.6 $\pm$  2.1    &    7.7 $\times 10^{-16}$ &  0.05 \\   %2_19
 5\dotfill &   716 $\pm$   86 &     593 $\pm$   58 &     94 $\pm$   18 &     96 $\pm$   18 &   52.8 $\pm$ 16.8    &    5.0 $\times 10^{-17}$ &  0.00 \\   %2_31
 6\dotfill &   \nodata        &      21 $\pm$    2 &     96 $\pm$    8 &     76 $\pm$    6 &    0.0 $\pm$  3.2    &    2.4 $\times 10^{-16}$ &  0.00 \\   %2_28
 7\dotfill &   \nodata        &      99 $\pm$    5 &     70 $\pm$    7 &     57 $\pm$    5 &   12.2 $\pm$  5.1    &    5.1 $\times 10^{-16}$ &  0.11 \\   %2_29
 8\dotfill &   265 $\pm$   26 &     248 $\pm$   12 &     66 $\pm$    7 &     57 $\pm$    7 &    0.0 $\pm$  4.4    &    3.7 $\times 10^{-16}$ &  0.32 \\   %2_33
 9\dotfill &   \nodata        &      11 $\pm$    1 &     90 $\pm$    7 &     80 $\pm$    7 &    0.0 $\pm$  4.9    &    2.6 $\times 10^{-16}$ &  0.00 \\   %2_32
10\dotfill &   338 $\pm$   21 &     121 $\pm$    6 &     50 $\pm$    5 &     23 $\pm$    6 &           \nodata    &  4.4 $\times 10^{-16}$ &  0.00 \\   %1_24
11\dotfill &   509 $\pm$   78 &      36 $\pm$    5 &     51 $\pm$   32 &     81 $\pm$   29 &           \nodata    &  6.5 $\times 10^{-17}$ &  0.00 \\   %1_17
12\dotfill &   639 $\pm$  109 &      57 $\pm$    7 &     88 $\pm$   15 &      0 $\pm$   13 &           \nodata    &  2.8 $\times 10^{-16}$ &  0.39 \\   %1_23
13\dotfill &   214 $\pm$   19 &     108 $\pm$    6 &    107 $\pm$   10 &     57 $\pm$    7 &           \nodata    &  1.8 $\times 10^{-16}$ &  0.20 \\   %3_35
14\dotfill &   113 $\pm$    7 &      14 $\pm$    1 &     96 $\pm$    5 &     29 $\pm$    3 &           \nodata    &  1.4 $\times 10^{-15}$ &  0.14 \\   %3_34
15\dotfill &   440 $\pm$   59 &      73 $\pm$    5 &     58 $\pm$   15 &     30 $\pm$   12 &           \nodata    &  7.3 $\times 10^{-17}$ &  0.00 \\   %1_09
16\dotfill &    94 $\pm$    6 &       0 $\pm$    1 &     85 $\pm$    6 &     42 $\pm$    5 &           \nodata    &  2.4 $\times 10^{-16}$ &  0.00 \\   %3_33
17\dotfill &   211 $\pm$   12 &      44 $\pm$    2 &    138 $\pm$    9 &     84 $\pm$    6 &           \nodata    &  2.9 $\times 10^{-16}$ &  0.00 \\   %3_32
18\dotfill &   206 $\pm$   25 &      22 $\pm$    4 &    114 $\pm$   14 &     60 $\pm$    7 &           \nodata    &  1.5 $\times 10^{-16}$ &  0.15 \\   %1_28
19\dotfill &   216 $\pm$   22 &      58 $\pm$    4 &    110 $\pm$    9 &     44 $\pm$    4 &    0.0 $\pm$  2.7    &    4.0 $\times 10^{-16}$ &  0.35 \\   %1_27
20\dotfill &   348 $\pm$   18 &     287 $\pm$   13 &     37 $\pm$    3 &     30 $\pm$    2 &   10.8 $\pm$  1.9    &    1.6 $\times 10^{-15}$ &  0.00 \\   %4_19
21\dotfill &   184 $\pm$   17 &      42 $\pm$    3 &    118 $\pm$   12 &     39 $\pm$    5 &    0.0 $\pm$  3.8    &    2.0 $\times 10^{-16}$ &  0.00 \\   %1_30
22\dotfill &   185 $\pm$   20 &      17 $\pm$    3 &    126 $\pm$   10 &     83 $\pm$    6 &    0.0 $\pm$  2.8    &    7.9 $\times 10^{-16}$ &  0.47 \\   %1_31
23\dotfill &   128 $\pm$    7 &      50 $\pm$    2 &     92 $\pm$    6 &     40 $\pm$    3 &           \nodata    &  6.2 $\times 10^{-16}$ &  0.04 \\   %3_30
24\dotfill &    76 $\pm$    7 &       0 $\pm$    1 &     79 $\pm$    6 &     50 $\pm$    5 &           \nodata    &  4.0 $\times 10^{-16}$ &  0.12 \\   %3_29
25\dotfill &   485 $\pm$   38 &     123 $\pm$    6 &     49 $\pm$    8 &     91 $\pm$   11 &    0.0 $\pm$  7.8    &    4.6 $\times 10^{-16}$ &  0.11 \\   %4_11
26\dotfill &   464 $\pm$   35 &      67 $\pm$    4 &     46 $\pm$    5 &     43 $\pm$    5 &    0.0 $\pm$  2.9    &    3.1 $\times 10^{-16}$ &  0.00 \\   %1_01
27\dotfill &   461 $\pm$   29 &     302 $\pm$   14 &     47 $\pm$    5 &     65 $\pm$    6 &    0.0 $\pm$  4.7    &    6.1 $\times 10^{-16}$ &  0.05 \\   %4_12
28\dotfill &   381 $\pm$   77 &       0 $\pm$    6 &     49 $\pm$   25 &      0 $\pm$   37 &    0.0 $\pm$ 27.8    &    5.1 $\times 10^{-17}$ &  0.00 \\   %4_13
29\dotfill &   254 $\pm$   14 &     192 $\pm$    9 &     59 $\pm$    3 &     30 $\pm$    1 &    9.6 $\pm$  0.6    &    5.6 $\times 10^{-15}$ &  0.00 \\   %3_27
30\dotfill &   382 $\pm$   22 &     110 $\pm$    5 &     55 $\pm$    5 &     38 $\pm$    5 &    0.0 $\pm$  4.7    &    7.5 $\times 10^{-16}$ &  0.12 \\   %4_18
31\dotfill &   389 $\pm$   21 &     105 $\pm$    5 &     46 $\pm$    3 &     44 $\pm$    2 &    6.6 $\pm$  1.2    &    7.2 $\times 10^{-16}$ &  0.00 \\   %1_02
32\dotfill &   716 $\pm$  104 &       0 $\pm$    4 &     57 $\pm$   17 &    111 $\pm$   27 &    0.0 $\pm$  2.0    &    1.0 $\times 10^{-16}$ &  0.00 \\   %4_17
33\dotfill &   342 $\pm$   29 &     112 $\pm$    6 &     54 $\pm$   12 &    103 $\pm$   11 &    0.0 $\pm$  5.7    &    1.9 $\times 10^{-16}$ &  0.00 \\   %1_03
34\dotfill &   213 $\pm$   14 &      19 $\pm$    1 &    121 $\pm$    7 &     62 $\pm$    4 &    0.0 $\pm$  1.4    &    1.3 $\times 10^{-15}$ &  0.23 \\   %1_32
35\dotfill &   101 $\pm$    8 &       0 $\pm$    1 &     99 $\pm$    7 &     41 $\pm$    4 &           \nodata    &  2.8 $\times 10^{-16}$ &  0.05 \\   %3_24
36\dotfill &  1071 $\pm$   82 &     269 $\pm$   13 &    324 $\pm$   23 &    224 $\pm$   12 &           \nodata    &  3.0 $\times 10^{-16}$ &  0.33 \\   %3_23
37\dotfill &   \nodata        &      23 $\pm$    3 &     85 $\pm$   18 &    110 $\pm$   22 &    0.0 $\pm$ 12.9    &    8.7 $\times 10^{-17}$ &  0.05 \\   %3_22
38\dotfill &    88 $\pm$   21 &       0 $\pm$    3 &    108 $\pm$   13 &     82 $\pm$    9 &    0.0 $\pm$  5.2    &    2.2 $\times 10^{-16}$ &  0.42 \\   %1_26
39\dotfill &   257 $\pm$   13 &     258 $\pm$   12 &     37 $\pm$    2 &     32 $\pm$    1 &    9.4 $\pm$  0.7    &    1.3 $\times 10^{-15}$ &  0.00 \\   %3_19
40\dotfill &   352 $\pm$   19 &     192 $\pm$    9 &     43 $\pm$    3 &     38 $\pm$    2 &   12.9 $\pm$  1.8    &    3.8 $\times 10^{-16}$ &  0.09 \\   %3_18
41\dotfill &   159 $\pm$   29 &     340 $\pm$   26 &     53 $\pm$   11 &     65 $\pm$   14 &    0.0 $\pm$  9.8    &    7.3 $\times 10^{-17}$ &  0.00 \\   %3_17
42\dotfill &   198 $\pm$   10 &      60 $\pm$    3 &    121 $\pm$    5 &     36 $\pm$    1 &           \nodata    &  2.4 $\times 10^{-15}$ &  0.31 \\   %3_16
43\dotfill &   336 $\pm$   18 &     274 $\pm$   12 &     39 $\pm$    3 &     30 $\pm$    2 &    8.5 $\pm$  1.6    &    5.4 $\times 10^{-16}$ &  0.00 \\   %3_15
44\dotfill &   193 $\pm$   15 &      13 $\pm$    2 &    127 $\pm$   12 &     58 $\pm$    7 &           \nodata    &  3.1 $\times 10^{-16}$ &  0.00 \\   %3_14
45\dotfill &   539 $\pm$   37 &      25 $\pm$    2 &     59 $\pm$    3 &     65 $\pm$    4 &    0.0 $\pm$  2.0    &    5.5 $\times 10^{-16}$ &  0.60 \\   %3_13
46\dotfill &    97 $\pm$    5 &       6 $\pm$    1 &    111 $\pm$    6 &     51 $\pm$    2 &           \nodata    &  1.3 $\times 10^{-15}$ &  0.05 \\   %3_12
47\dotfill &   280 $\pm$   36 &      43 $\pm$    4 &    114 $\pm$   14 &     70 $\pm$   11 &    0.0 $\pm$  8.7    &    8.7 $\times 10^{-17}$ &  0.09 \\   %3_11
48\dotfill &   \nodata        &     121 $\pm$    6 &     62 $\pm$    6 &     51 $\pm$    6 &    0.0 $\pm$  5.3    &    2.7 $\times 10^{-16}$ &  0.00 \\   %3_06
49\dotfill &   399 $\pm$   29 &      33 $\pm$    2 &    105 $\pm$    8 &     64 $\pm$    6 &           \nodata    &  3.0 $\times 10^{-16}$ &  0.19 \\[0mm]   %3_02
\enddata
\tablecomments{Line fluxes in columns (2) and (3) are in units of H$\beta$\,=\,100, those in columns (4)--(6) are in units of H$\alpha$\,=\,286. The extinction correction has been applied to F(\hbeta) and the error on c(\hbeta) has been propagated to the line fluxes in columns (2)--(6).}
\end{deluxetable*}

%% file: tab3.tex
\begin{deluxetable}{ccccc}
\tabletypesize{\scriptsize}
\tablecolumns{5}
%\tablewidth{300pt}
\tablewidth{0pt}
\tablecaption{Direct method abundances\label{auroral}}

\tablehead{
\colhead{\phantom{}ID\phantom{}}	     &
\colhead{\oiii}	 &
\colhead{T$_e$}  &
\colhead{12\,+\,log(O/H)}	 &
\colhead{log(N/O)}	 \\[0.5mm]
\colhead{}       &
\colhead{4363}   &
\colhead{(K)}   &
\colhead{}   &
\colhead{} \\[1mm]
\colhead{(1)}	&
\colhead{(2)}	&
\colhead{(3)}	&
\colhead{(4)}	&
\colhead{(5)}	}
\startdata
\\[-1mm]
20\dotfill	&	2.8 $\pm$ 0.3   & 11550 $\pm$ 660 & 8.17 $\pm$ 0.07 & $-1.18$ $\pm$ 0.08 \\  %4_19
29\dotfill	&	0.7 $\pm$ 0.1	&	8730 $\pm$ 470 & 8.41 $\pm$ 0.08 & $-0.98$ $\pm$ 0.09 \\  %3_27
39\dotfill	&	2.7 $\pm$ 0.3	&	11810 $\pm$ 640 & 8.05 $\pm$ 0.07 & $-1.03$ $\pm$ 0.08 \\  %3_19
43\dotfill	&	2.9 $\pm$ 0.6	&	11895 $\pm$ 1080 & 8.12 $\pm$ 0.12 & $-1.14$ $\pm$ 0.14 \\  %3_15

\enddata
\tablecomments{The reddening-corrected intensity of \oiii\lin4363 is in units of H$\beta$\,=\,100.\bigskip}
\end{deluxetable}

%% file: tab4.tex
\begin{deluxetable}{ccccc}
\tabletypesize{\scriptsize}
\tablecolumns{5}
%\tablewidth{300pt}
\tablewidth{0pt}
\tablecaption{Emission-line galaxies\label{galaxies}}

\tablehead{
\colhead{\phantom{aaaaa}ID\phantom{aaaaa}}	     &
\colhead{R.A.}	 &
\colhead{Decl.}  &
\colhead{$z$}	 &
\colhead{Field}	 \\[0.5mm]
\colhead{}       &
\colhead{(J2000.0)}   &
\colhead{(J2000.0)}   &
\colhead{}   &
\colhead{} \\[1mm]
\colhead{(1)}	&
\colhead{(2)}	&
\colhead{(3)}	&
\colhead{(4)}	&
\colhead{(5)}	}
\startdata
\\[-1mm]
1\dotfill  & 13 36 07.91  &  $-$29 53 17.9  &   0.517 &	 2     \\      % F2-7
2\dotfill  & 13 36 08.20  &  $-$29 55 10.4  &   0.160 &	 2     \\      % F2-10
3\dotfill  & 13 36 11.19  &  $-$29 54 24.2  &   0.528 &	 2     \\      % F2-12
4\dotfill  & 13 36 14.76  &  $-$29 53 56.6  &   0.268 &	 2     \\      % F2-14
5\dotfill  & 13 36 28.20  &  $-$29 53 50.5  &   0.305 &	 2     \\      % F2-26
6\dotfill  & 13 36 51.27  &  $-$29 45 57.1  &   0.426 &	 1     \\      % F1-20
7\dotfill  & 13 36 59.76  &  $-$30 00 57.5  &   0.306 &	 3     \\      % F3-25
8\dotfill  & 13 37 00.46  &  $-$29 41 59.0  &   0.317 &	 1     \\      % F1-6
9\dotfill  & 13 37 13.44  &  $-$29 57 43.6  &   0.306 &	 3     \\      % F3-10
\enddata
\tablecomments{Units of right ascension are hours, minutes and seconds, and units of declination
are degrees, arcminutes and arcseconds. Col.~(1): \hii\/ region identification. Col.~(2): Right
Ascension. Col.~(3): Declination. Col.~(4): Redshift.
Col.~(5): FORS field.\\ \\ \\ \\ \\ \\}
\end{deluxetable}